\documentclass[10pt,sigconf,nonacm]{acmart}

\usepackage[ruled]{algorithm2e}
\usepackage{balance}       
\usepackage{subcaption}
\usepackage{graphicx}
\usepackage{varwidth}
\usepackage{url}
\usepackage{tabularx}
\usepackage{xcolor}
\usepackage[utf8]{inputenc}

\usepackage{amsmath}
\usepackage{algorithmic}
\usepackage{booktabs}
\usepackage{enumitem}
\usepackage{xspace}
\usepackage{listings}
\usepackage{pifont}
\usepackage{bbding}
\usepackage{wasysym}
\usepackage{phonetic}
\usepackage{tipa}
\usepackage{interval}
\newcommand{\cmark}{\ding{51}}
\newcommand{\xmark}{\ding{55}}
\newcommand{\pmark}{\raisebox{0.2ex}{\footnotesize$\triangle$}}

\definecolor{hyperlink}{RGB}{65,76,135}

\newcommand{\ours}{\textsc{Ultrasound Watermark}\xspace}

\begin{document}

\title{\ours: Real-time Acoustic Watermarking against Voice Scams on Smartphones}

\author{Renzhi Hao}
\email{horange1024@outlook.com}
\affiliation{%
  \institution{Bloomberg L.P.}
  \city{New York}
  \state{New York}
  \country{USA}
}

\author{Shirui Cao}
\email{shiruicao@cs.umass.edu}
\affiliation{%
  \institution{University of Massachusetts Amherst}
  \city{Amherst}
  \state{Massachusetts}
  \country{USA}
}

\author{Yizhu Wen}
\email{yizwen@iu.edu}
\affiliation{%
  \institution{Indiana University Bloomington}
  \city{Bloomington}
  \state{Indiana}
  \country{USA}
}

\author{Rishav Gupta}
\email{rishavg1@umbc.edu}
\affiliation{%
  \institution{University of Maryland, Baltimore County}
  \city{Baltimore}
  \state{Maryland}
  \country{USA}
}

\author{Dharani Nadendla}
\email{dharann1@umbc.edu}
\affiliation{%
  \institution{University of Maryland, Baltimore County}
  \city{Baltimore}
  \state{Maryland}
  \country{USA}
}

\author{Mehran Kafai}
\email{mkafai@amazon.com}
\affiliation{%
  \institution{Amazon}
  \city{San Diego}
  \state{California}
  \country{USA}
}
\author{Hanqing Guo}
\email{guohan@iu.edu}
\affiliation{%
  \institution{Indiana University Bloomington}
  \city{Bloomington}
  \state{Indiana}
  \country{USA}
}

\author{Dong Li}
\email{dli@umbc.edu}
\affiliation{%
  \institution{University of Maryland, Baltimore County}
  \city{Baltimore}
  \state{Maryland}
  \country{USA}
}

\begin{abstract}
Recent advances in text-to-speech (TTS), voice conversion (VC), and AI-driven speech synthesis enable adversaries to convincingly impersonate trusted individuals during phone calls. Existing defenses often rely on offline analysis, degrade under telephony distortions, or remain vulnerable to adaptive attacks. To address these limitations, we present \ours, a real-time, smartphone-native framework that protects live calls by binding speech to the caller's physical articulatory motion through acoustic watermarking. During a call, the smartphone emits ultrasonic signals and captures reflections from facial articulatory movements using only its built-in speaker and microphone. \ours extracts motion features from these reflections and fuses them with outgoing speech to generate a dynamic watermark. Because the ultrasonic measurements are physically and temporally coupled with speech production, they are difficult to spoof without the underlying articulatory motion. A synchronized generator-detector architecture enforces cross-modal speech-motion consistency against replay, synthesized speech, and articulation-mimicking attacks. \ours operates continuously with less than 200 ms latency overhead and requires no additional hardware. We implement \ours on iOS and Android and evaluate it using over 2.5 hours of synchronized speech-ultrasound recordings from 22 speakers across eight smartphones and four real-world environments. Results demonstrate robustness to diverse attacks and telephony distortions while preserving perceptual audio quality.

\end{abstract} 
\maketitle


 \begin{figure}
    \centering
    \includegraphics[width=0.75\linewidth]{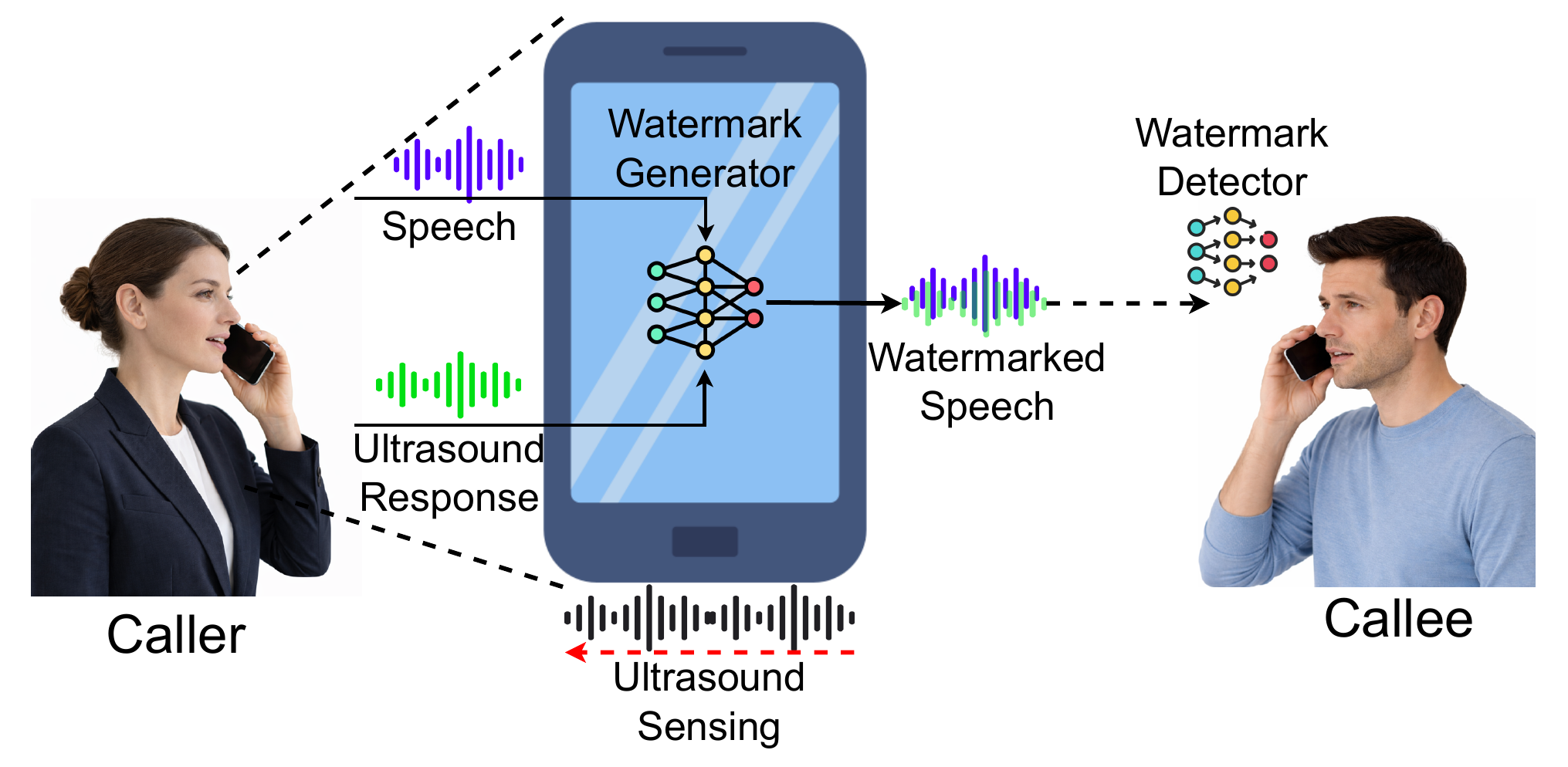}
     \caption{\ours usage scenario.}
     \label{fig:cover}
 \end{figure}  
\section{Introduction}\label{sec-introduction}

    

    

    
     

Voice is a natural form of human communication that carries both linguistic information and speaker-specific biometric cues. In mobile environments, it increasingly serves as an identifier for copyright protection, data access, and secure communication. However, advances in generative AI have introduced significant security threats. Modern text-to-speech (TTS) and voice conversion (VC) systems, including WaveNet- and GAN-based models~\cite{shen2018natural, kong2020hifi, qian2019autovc, kameoka2018stargan}, can synthesize voices that closely mimic real individuals with minimal training data. Such capabilities have fueled impersonation attacks such as voice phishing (``vishing''), which caused over \$2.95 billion in losses to U.S. consumers in 2024 alone~\cite{ftc2025sentinel, pappas2025hiddenthreat}. Meanwhile, humans often struggle to distinguish synthesized voice clones from genuine speech, with detection performance approaching chance levels~\cite{nature2025voice, figueiredo2024vishing}.


\noindent\textbf{Scenario.} 
Consider a scenario where a user receives a call that appears to come from someone they know. The caller may pose as a supervisor and request urgent actions like approving a payment. Because the voice sounds familiar, the callee may trust the call and act quickly. A successful attack can cause financial loss, data leakage, and long-term damage to trust and professional relationships. In this setting, the callee must decide in real time whether the voice is from a legitimate human or a synthetic impersonation.

\noindent\textbf{Existing Efforts.} As shown in Table~\ref{tab:comparison}, prior defenses fall short in this scenario. \emph{Deepfake Detectors}, such as ASVspoof-style systems~\cite{yamagishi2021asvspoof,add_2022,yi2023add2023secondaudio}, analyze audio content to identify artifacts introduced by synthesized speech. These methods are typically applied offline or on the server side. While they can defend against replay and synthesis attacks, acoustic artifacts they rely on are often suppressed or distorted in real phone calls by telephony codecs, echo, and streaming effects, making them poorly suited for detecting impersonation during live calls. \emph{Digital Watermarking} embeds imperceptible bits into legitimate speech that can later be extracted or detected for authentication~\cite{chen2023wavmark,roman2024proactive,zong2025audiomarknet,liu2025xattnmark,singh2024silentcipher}. However, because watermark embedding and verification are typically performed offline, these methods cannot verify in real time that speech originates from a legitimate speaker during a call. \emph{Liveness Detection} verifies that audio originates from a live human speaker and may require specialized sensors or explicit user interaction~\cite{guo2022supervoice,li2024piezobud}. However, liveness alone does not establish identity: an attacker may pass liveness detection by speaking into the phone while injecting or replaying synthesized speech to impersonate a trusted person, such as a family member or supervisor~\cite{zhang2017hearing,li2024piezobud,shahid2023my}. Finally, \emph{broader provenance initiatives and platform-level policies} support media origin tracking but do not cryptographically or physically bind the audio to a legitimate human speaker during an ongoing phone call~\cite{deepmind2024synthid,openai2024source}. Collectively, these limitations highlight the lack of a practical in-call defense that can operate under real-time conditions, establish liveness, and verify speech legitimacy.

\begin{table*}[t]
\centering
\footnotesize
\setlength{\tabcolsep}{4pt}
\begin{tabularx}{\linewidth}{lcccccccc}
\toprule
\textbf{Approach} &
\textbf{In-call} &
\textbf{No Extra} &
\textbf{Phone} &
\textbf{Replay} &
\textbf{AI Synthesis} &
\textbf{Live Human} &
\textbf{User Effort} &
\textbf{References} \\
 &  & \textbf{HW} & \textbf{Native} & \textbf{Attack} & \textbf{(TTS/VC)} & \textbf{Impersonation} &  &  \\
\midrule
Deepfake Detectors (ASVspoof, ML) 
& \pmark & \cmark & \pmark 
& \pmark 
& \cmark 
& \xmark 
& \pmark 
& \cite{yamagishi2021asvspoof,add_2022,yi2023add2023secondaudio} \\

Audio Watermarking 
& \xmark & \cmark & \pmark 
& \xmark 
& \pmark 
& \xmark 
& \cmark 
& \cite{chen2023wavmark,roman2024proactive,zong2025audiomarknet,liu2025xattnmark,singh2024silentcipher} \\

Liveness Detection (sensors, protocols) 
& \pmark & \textbf{Often} & \pmark 
& \cmark 
& \pmark 
& \xmark 
& \pmark 
& \cite{zhang2017hearing,li2024piezobud,shahid2023my} \\

Provenance (platform-level) 
& \xmark & \cmark & \cmark 
& \xmark 
& \pmark 
& \xmark 
& \cmark 
& \cite{deepmind2024synthid,openai2024source} \\

\midrule
\textbf{Our work: \ours} 
& \textbf{\cmark} 
& \textbf{\cmark} 
& \textbf{\cmark} 
& \textbf{\cmark} 
& \textbf{\cmark} 
& \textbf{\cmark} 
& \textbf{Low} 
& \textit{-} \\
\bottomrule
\end{tabularx}
\caption{Comparison of defenses against voice impersonation in live calls on commodity smartphones. \textit{In-call}: operates during a call; \textit{No Extra HW}: requires no additional hardware; \textit{Phone-native}: uses stock phone hardware; \textit{Replay}, \textit{AI Synthesis}, and \textit{Live Human}: robustness to recorded speech, synthesized/converted speech, and natural human impersonation, respectively; \textit{User Effort}: interaction burden. \emph{Symbols:} \cmark, \xmark, and \pmark\ denote supported, unsupported, and partial/conditional support, respectively.}
\label{tab:comparison}
\end{table*}

To defend against in-call impersonation attacks, we propose \ours, a real-time, smartphone-native caller verification system that binds speech to the caller's physical articulation through acoustic watermarking. \ours consists of a watermark generator on the caller's phone and a detector on the callee's phone. As shown in Fig.~\ref{fig:cover}, the caller's phone continuously emits ultrasound and captures reflections from orofacial motion (i.e., the mouth, jaw, and cheeks) using its built-in speaker--microphone pair. These signals are fused with outgoing speech to generate a dynamic watermark, which is imperceptibly embedded into the speech and transmitted to the callee. The detector verifies the caller's legitimacy through three layers: (i) \textbf{Protocol Compliance}, ensuring the received signal follows the expected watermarking pipeline and supports valid decoding; (ii) \textbf{Live Speech Presence}, detecting ultrasound reflections from articulatory motion to confirm real-time speech production near the caller's phone; and (iii) \textbf{Speech-Motion Consistency}, verifying that the speech matches the observed articulatory motion.

\noindent\textbf{Challenges.}
Building such a system is non-trivial due to four major challenges. 
(\textbf{C1: Ultrasound Feature Extraction and Fusion.}) The system must reliably extract articulatory motion cues from ultrasonic reflections and fuse them with audible speech in real time, using only commodity smartphone hardware. These features must be compact, temporally aligned with speech, and expressive enough to capture fine-grained articulatory motion that cannot be inferred from speech alone. 
(\textbf{C2: Adversarial Realism.}) The watermark must remain effective under attacks, including replay, synthesized speech, and live human impersonation with mismatched or loosely synchronized articulatory motion, as well as under real call distortions such as codec compression, echo, and noises. This requires enforcing causal alignment between speech and ultrasound rather than relying on either modality independently. 
(\textbf{C3: Real-time Performance.}) All processing must run continuously with low latency and bounded computation, generating and verifying watermarks frame by frame while keeping end-to-end delay imperceptible.
(\textbf{C4: Cross-Platform Deployment.}) Finally, the system must run entirely on stock iOS and Android smartphones without additional hardware or user interaction, despite platform-specific constraints on signal processing, on-device model execution, and audio pipeline integration.

To address these challenges, we design \ours as a streaming, dual-modality system with three key designs: (i) a lightweight ultrasound processing pipeline extracts phase- and amplitude-based motion features and fuses them with speech through a cross-modal encoder-decoder, generating watermarks tied to live articulatory motion; (ii) a synchronized generator--detector architecture enforces fine-grained causal alignment between speech and motion, preventing replayed, synthesized, or articulation-mimicking speech from producing consistent cross-modal correspondence; and (iii) a sliding-window streaming design enables low-latency operation, while device-aware calibration, codec-aware optimization, and redundancy improve robustness across devices, environments, and call dynamics. Together, these designs enable physiologically grounded, real-time speech verification on smartphones while preserving call quality. We make the following contributions:
\begin{figure*}
    \centering
    \includegraphics[width=1\linewidth]{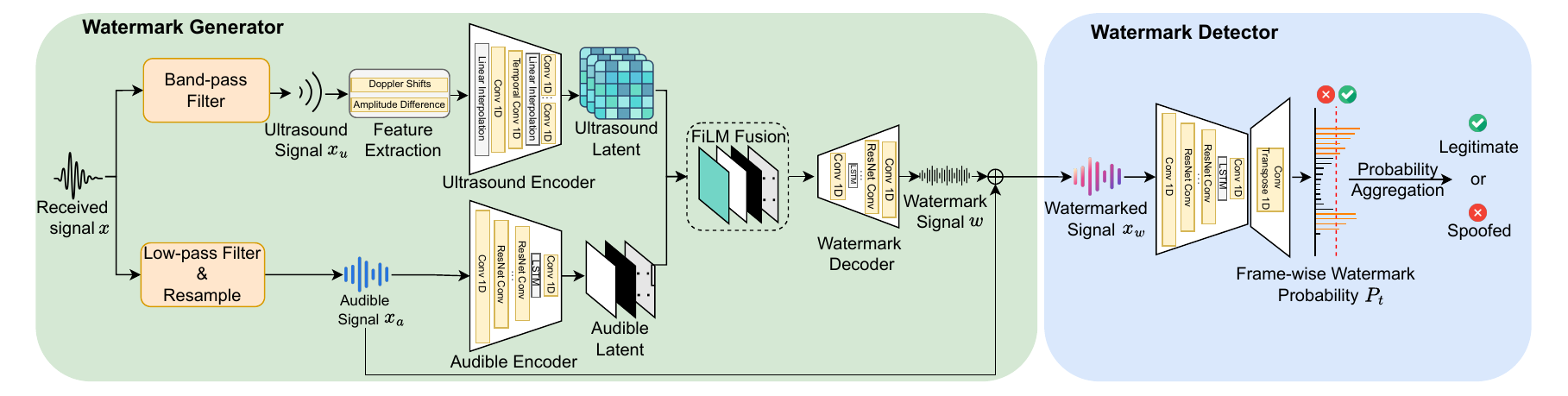}
    \caption{Overview of our proposed ultrasound watermark framework.}
    \label{fig:watermark_design}
\end{figure*}
\begin{itemize}[leftmargin=*]
\item \textbf{Real-time Ultrasound Watermarking Framework:}
We propose \ours, a real-time acoustic watermarking framework that binds live speech to physical articulatory motion using smartphone ultrasound. By jointly encoding speech and ultrasound reflections into a frame-wise watermark and enforcing their temporal consistency at the detector side, \ours provides real-time verification against replay, synthesis, and articulation-mimicking attacks.
\item \textbf{Comprehensive Evaluation:}
Under an IRB-approved protocol, we collect over 2.5 hours of synchronized speech--ultrasound recordings from 22 speakers using eight commodity smartphones across four real-world environments. We evaluate \ours against audio-only, random-ultrasound, speech-synchronized ultrasound, and adaptive adversarial attacks, considering both replayed and synthesized speech. We further compare against recent audio-only liveness detection methods and evaluate robustness under common telephony distortions and codecs. Results across detection, attack success, latency, and speech quality metrics demonstrate robust protection with limited impact on perceptual audio quality.
\item \textbf{Cross-Platform Mobile System:}
We implement \ours as a complete on-device prototype for both iOS and Android, requiring no extra hardware. The prototype performs real-time watermark generation and detection during live calls under practical attack scenarios. A real-time system demo is available \href{https://ultrasoundwatermark.github.io}{\color{blue}{here}}. Source code and dataset will be released upon acceptance.
\end{itemize}

\section{Threat Model}\label{sec-threat}

\noindent\textbf{User Setting.} The victim is a security-conscious callee who \emph{makes sensitive decisions based on perceived caller identity}. We assume the victim uses \ours for calls involving financial or sensitive information and expects the other party to use the same app. This setting is practical because secure calling is commonly provided via application-level services requiring both endpoints to adopt the same platform (e.g., WhatsApp~\cite{whatsapp-requirement} and Apple FaceTime~\cite{apple-facetime-requirements}).

\noindent\textbf{Attacker's Goal.} The attacker aims to impersonate a trusted speaker (e.g., a family member or supervisor) and induce the callee to take a security-sensitive action. Specifically, the attacker seeks to make the callee-side detector accept attacker-controlled speech as carrying a valid watermark, causing the call to appear legitimate.

\noindent\textbf{Attacker Capabilities.} We assume the attacker can place real-time calls using smartphones and obtain prior voice samples of the target speaker (e.g., from social media) for replay or speech synthesis. The attacker may run our app on their own device but cannot remotely access the target call's live ultrasonic sensing path or directly reproduce the target speaker's ultrasound response coupled to the physical acoustic path. Thus, replayed or synthesized speech may mimic the target's voice but does not inherently reproduce the synchronized speech--ultrasound coupling of live articulation. We consider four increasingly capable attack scenarios:
\begin{itemize}[leftmargin=*]
\item \textbf{Scenario 1: Audio-only Spoofing.} The attacker calls the victim without enabling ultrasound watermarking and controls only the speech. We consider two attacks: (i) \textit{Replay}, where previously recorded target speech is replayed during the call~\cite{wang2020asvspoof, Chen2018rnn}; and (ii) \textit{Synthesis}, where the target's voice is generated using TTS~\cite{casanova2022yourtts} or VC~\cite{li2023freevc, li2022freevchighqualitytextfreeoneshot}.

\item \textbf{Scenario 2: Random Ultrasound Spoofing.} The attacker is aware of \ours and can run the app on their own device but cannot generate ultrasound traces correctly coupled to the spoofed speech. While replaying or synthesizing the target's speech, the attacker produces arbitrary, unrelated mouth movements near the phone to induce ultrasonic reflections.

\item \textbf{Scenario 3: Speech-synchronized Ultrasound Spoofing.} The attacker adopts a stronger strategy by synchronizing their mouth movements with replayed or synthesized speech. By mimicking the corresponding articulation near the device, the attacker attempts to produce ultrasound reflections consistent with the spoofed speech and deceive \ours into accepting the combined signals as originating from a legitimate caller.

\item \textbf{Scenario 4: Adaptive Adversarial Attacks.} The attacker has system knowledge and adapts spoofed signals to the generator--detector framework. The attacker may use detector queries, architecture knowledge, or surrogate models to craft perturbations that cause spoofed speech to be accepted. They may also modify replayed or synthesized speech before watermark generation, aiming to have it accepted as originating from a legitimate caller.
\end{itemize}

\section{System Design}\label{sec-system}

\noindent\textbf{Overview:}
We propose \textsc{Ultrasound Watermark}, a dual-modality framework that embeds articulatory cues from ultrasonic reflections into speech for real-time legitimate caller verification on smartpones. As shown in Fig.~\ref{fig:watermark_design}, the system consists of a \emph{Watermark Generator} and a \emph{Watermark Detector}. The generator jointly encodes audible speech and ultrasonic signals, fusing speech latents with the ultrasound feature map that captures instantaneous phase and amplitude variations. This design grounds the watermark in live articulatory motion rather than acoustic replay. The detector performs lightweight on-device inference, estimating frame-level watermark confidence and aggregating it over time for integrity verification. All components are optimized for low latency, robustness, and smartphone deployment, 
enabling continuous, imperceptible legitimate caller verification.

\begin{figure}
    \centering
    \begin{subfigure}[b]{0.49\linewidth}
        \centering
        \includegraphics[width=\linewidth]{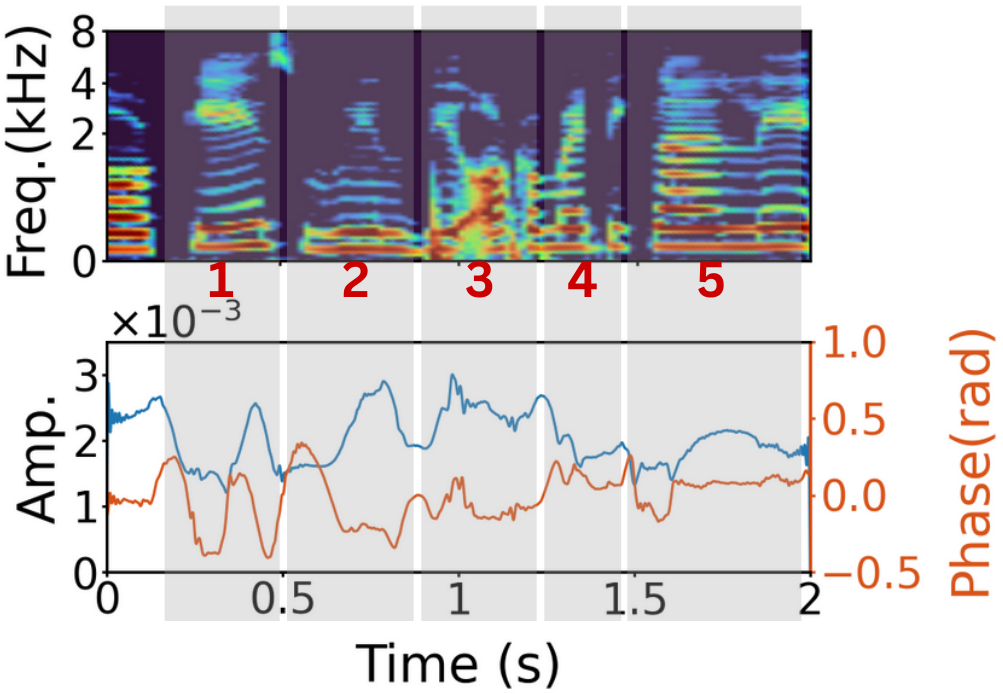}
        \caption{Speech vs. ultrasound.}
        \label{fig:ultrasound_feature_normal}
    \end{subfigure}
    \hfill
    \begin{subfigure}[b]{0.49\linewidth}
        \centering  
        \includegraphics[width=\linewidth]{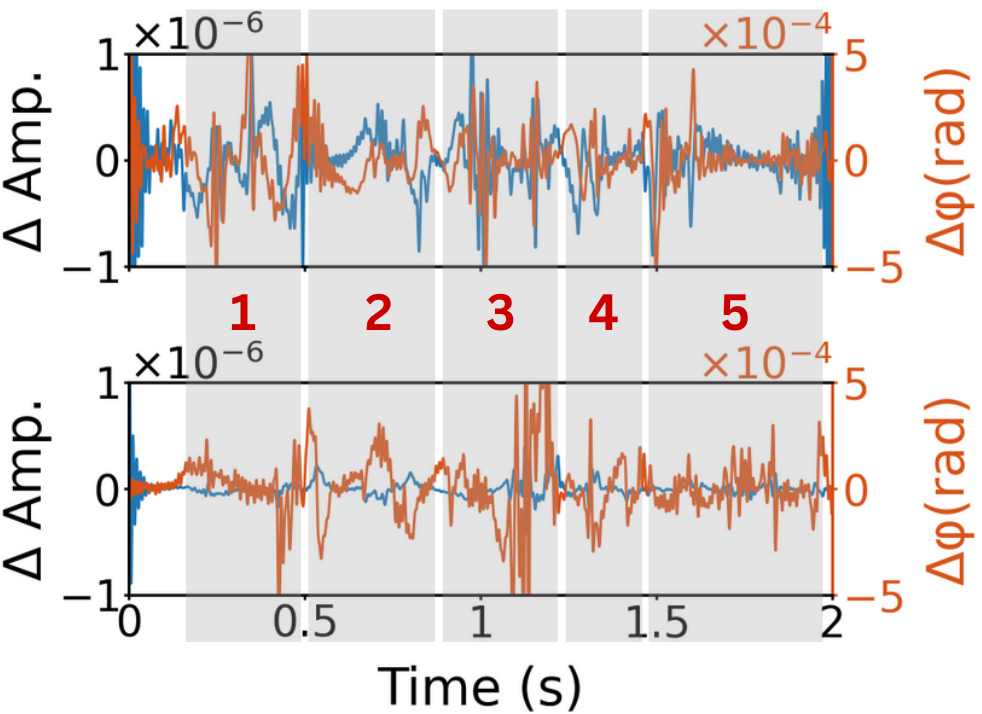}
        \caption{Legitimate vs. spoofed.}
        \label{fig:ultrasound_feature_mimic}
    \end{subfigure}
    \caption{Ultrasound features during speech. Shaded regions denote spoken words: 1: ``Your'', 2: ``case'', 3: ``will'', 4: ``be'', 5: ``automatically''. (a) Speech Mel spectrogram with ultrasonic amplitude and phase features. (b) Phase and amplitude differences for legitimate (Top) and speech-synchronized spoofed ultrasound (Bottom).}
    \label{fig:ultrasound_features}
\end{figure}

\subsection{Feature Extraction and Fusion}
\label{sec:resolve_c1}
We propose a lightweight sensing design that extracts a minimal set of motion cues that are causally tied to speech production, difficult to spoof via replay or synthesis, and efficient for continuous in-call operation.

\noindent\textbf{Design Rationale.} Our design is guided by three insights. First, articulatory motion occurs at millimeter and millisecond scales that low-rate frame-based features fail to capture. Continuous-wave ultrasound enables fine-grained sample-level sensing aligned with phonetic events~\cite{sun2021ultrase,zhang2021sensing}. Second, spoof resistance requires physically grounded features. While absolute spectral patterns can be replayed or synthesized, temporal derivatives of phase and amplitude induced by live motion are harder to forge. Third, mobile constraints rule out heavyweight and non-causal processing, requiring an incremental pipeline with lightweight signal processing.

During a call, the caller's smartphone continuously emits a multi-carrier ultrasonic signal~\cite{zhang2021sensing} with base frequency $f_0 = 20~\text{kHz}$ and six equally spaced carriers ($\Delta f = 300~\text{Hz}$):
$
u(t) = \sum_{i=0}^{5} \cos(2\pi (f_0+i\times\Delta f) t),
$
where $f_0+i\times\Delta f$ is the $i$-th carrier frequency. Meanwhile, the built-in microphone records $x(t)$ at 48~kHz, containing both ultrasonic reflections $x_u(t)$ and audible speech $x_a(t)$. Both originate from the same articulatory motion, providing natural temporal alignment without external synchronization.

\noindent\textbf{Ultrasound Feature Extraction.}
We isolate each ultrasonic carrier using lightweight finite impulse response (FIR) band-pass filters and apply cosine--sine (I/Q) mixing to demodulate the continuous-wave signal to baseband. This enables sample-level tracking of amplitude and phase to capture fine-grained articulatory motion aligned with speech. For each carrier $x_{u_i}(t)$, the in-phase and quadrature components are:

\begin{equation}
I_i(t) = x_{u_i}(t)\cos(2\pi f_i t),
\quad
Q_i(t) = -x_{u_i}(t)\sin(2\pi f_i t).
\end{equation}

After applying a moving average to suppress high-frequency artifacts, we recover the phase $\phi_i(t)$ and amplitude $A_i(t)$:
\begin{equation}
\phi_i(t) = \operatorname{atan}(Q_i(t), I_i(t)),
\quad
A_i(t) = \sqrt{I_i^2(t) + Q_i^2(t)}.
\end{equation}

Fig.~\ref{fig:ultrasound_feature_normal} shows the speech Mel spectrogram and phase and amplitude of the first carrier, whose variations reflect articulatory motion. To improve robustness, we compute phase and amplitude differences. Phase differences capture articulator velocity through the Doppler effect, while amplitude differences capture changes in reflector distance and orientation. Fig.~\ref{fig:ultrasound_feature_mimic} shows that these features distinguish legitimate from speech-synchronized spoofed ultrasound while suppressing static channel effects. Concatenating them across carriers yields a 12-channel feature map. A compact multi-branch temporal network captures articulatory patterns at different scales, followed by dilated convolutions for longer-range dependencies. A final projection produces an ultrasonic latent sequence $\mathbf{h}_u$ aligned with the audio latent space.

\noindent\textbf{Audible Speech Feature Extraction and Encoding.}
In parallel, the received signal is low-pass filtered and resampled from 48~kHz to 16~kHz using FIR filtering and linear interpolation, aligning the audio with telephony codecs while minimizing perceptual distortion. The audio pipeline is intentionally simpler than the ultrasound pipeline, as ultrasound provides the primary liveness signal while speech serves as the perceptual carrier. The resampled audio is encoded by a strided residual convolutional network inspired by AudioSeal~\cite{copet2023simple} and Encodec~\cite{defossez2022high}, which reduces temporal resolution while increasing channel dimensionality to produce a compact latent representation $\mathbf{h}_a$ that preserves phonetic and prosodic structure for watermark insertion.

\noindent\textbf{Cross-modal Fusion via Ultrasound-Guided FiLM.} To inject ultrasound context while preserving speech as the dominant carrier, we apply Feature-Wise Linear Modulation (FiLM) to the audio latent using the ultrasound sequence. Given $\mathbf{h}_u$ and $\mathbf{h}_a$, two temporal $1\times1$ convolutions $g_{\text{scale}}(\cdot)$ and $g_{\text{shift}}(\cdot)$ generate scale and shift fields:
\begin{equation}
\gamma = \sigma(g_{\text{scale}}(\mathbf{h}_u)), \quad
\beta = g_{\text{shift}}(\mathbf{h}_u),
\end{equation}
where $\sigma(\cdot)$ denotes the element-wise sigmoid function. The fused latent representation $\mathbf{z}$ is then computed as:
\begin{equation}
\mathbf{z} = (1 + \gamma) \odot \mathbf{h}_a + \beta,
\end{equation}
where $\odot$ denotes the Hadamard product.

\noindent\textbf{Watermark Decoder.}
The fused latent representation is decoded by a lightweight transposed convolutional network to reconstruct the audio carrier and generate the watermark. 




\subsection{Adversarially-aligned Training}

We design an adversarially-aligned training framework that forces the generator--detector pair to learn the synchrony between speech and ultrasonic articulatory motion.

\begin{figure}[t]
    \centering
    \includegraphics[width=\linewidth]{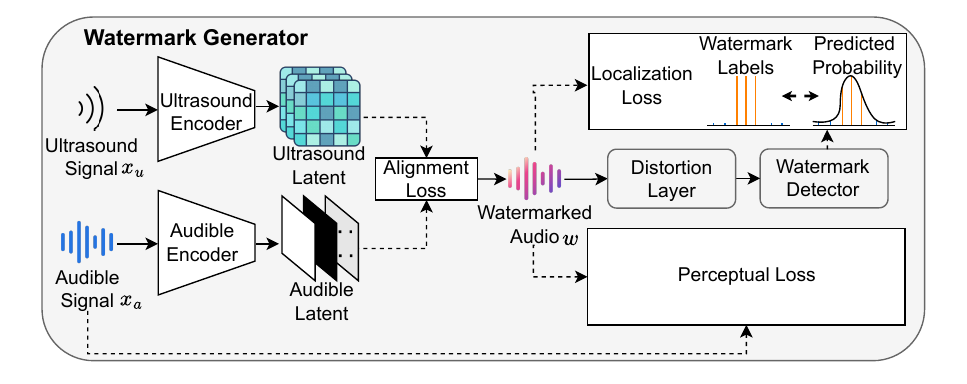}
    \caption{Generator-detector training pipeline.}
    \label{fig:training_pipeline}
\end{figure}

\noindent\textbf{Overview.}
Fig.~\ref{fig:training_pipeline} summarizes our approach. Paired audible speech and ultrasound signals are first routed through the watermark generator, which produces a watermarked audio. A synchronized distortion layer applies codec compression, echo, and reverberation to watermarked audios, ensuring that the model encounters realistic phone call degradations and spoof-like perturbations. The resulting watermarked audio is then processed by the Watermark Detector, which outputs frame level watermark probabilities. During training, speech is treated as a multimodal event in which acoustic sound waves and the sensed ultrasound are in lockstep. The model is trained so that only this co-evolving pattern triggers watermark activation, while any disruption of synchrony caused by replay attacks, mismatched motion, silent mimicry, or channel distortions is explicitly penalized.


\noindent\textbf{Watermark Detector.}
The detector is trained to distinguish synchronized speech and ultrasound pairs from mismatched ones, and to output frame-level watermark probabilities. Specifically,
the detector uses an Encodec-style encoder to extract a compact latent representation. A learned upsampling module restores the temporal resolution. A $1 \times 1$ convolutional head produces two logits at each time step. A softmax converts them into frame-level probabilities. During training, aligned speech and ultrasound pairs yield high watermark probabilities, reflecting genuine articulatory motion coupled with speech. In contrast, spoofed samples break this coupling and produce low watermark probabilities. At inference time, the detector aggregates frame-level outputs into a global confidence score by computing the fraction of frames whose watermark probability exceeds the detection threshold.

\noindent\textbf{Joint Generator–Detector Optimization.}
We jointly train the watermark generator and detector in an end-to-end fashion. The watermark generator operates on both genuine audios (with aligned audio-ultrasound pairs) and spoofed audios (with misaligned pairs), while the detector learns to distinguish between them. This joint optimization ensures that the watermark encoding captures temporal synchrony information that the detector can reliably verify at inference time. The overall training loss is:
\begin{equation}
  \mathcal{L} =  \lambda_{\text{loc}} \, \mathcal{L}_{\text{loc}}
+ \lambda_{\text{align}} \, \mathcal{L}_{\text{align}}
+ \lambda_{\text{percept}} \, \mathcal{L}_{\text{percept}}.
\end{equation}

\begin{itemize}[leftmargin=*]
    \item \textbf{Localization Loss:} The localization loss \(\mathcal{L}_{\text{loc}}\) supervises the detector's frame-level predictions against ground-truth labels. During training, genuine and spoofed watermarked audio are jointly sampled and optimized in parallel, allowing the model to learn the cross-modal correspondence that characterizes genuine speech while discriminating it from mismatched pairings. Let \(\mathbf{p}^{+}\) and \(\mathbf{p}^{-}\) denote the detector's softmax outputs for genuine and spoofed samples, respectively. The loss is computed via negative log-likelihood (NLL):
\begin{equation}
\mathcal{L}_{\text{loc}} = \mathcal{L}_{\text{NLL}}(\log \mathbf{p}^{+}, y^{+}) + \mathcal{L}_{\text{NLL}}(\log \mathbf{p}^{-}, y^{-}),
\end{equation}
where \(y^{+}\) and \(y^{-}\) are the ground-truth labels: \(y^{+} = 1\) indicates genuine watermarked audio (accept), and \(y^{-} = 0\) indicates spoofed watermarked audio (reject).

    \item \textbf{Alignment Loss:} Unlike classification-based approaches, we adopt a contrastive objective over learned hidden representations. Let $\mathbf{h}_a$, $\mathbf{h}_u^+$, and $\mathbf{h}_u^-$ be the $\ell_2$-normalized hidden features produced by the watermark encoder. Specifically, they are extracted from audible speech, aligned ultrasound, and misaligned ultrasound, respectively. The alignment loss \(\mathcal{L}_{\text{align}}\) encourages synchronized audio-ultrasound pairs to have similar representations while pushing misaligned pairs apart:

\begin{equation}
\begin{aligned}
\mathcal{L}_{\text{align}} =
& \underbrace{(1 - \cos(\mathbf{h}_a, \mathbf{h}_u^+))}_{\text{alignment term}} \\
& + \underbrace{\max(0, \cos(\mathbf{h}_a, \mathbf{h}_u^-) - \delta)}_{\text{misalignment term}},
\end{aligned}
\end{equation}
    where $\cos(\cdot, \cdot)$ is cosine similarity, and $\delta$ is a margin hyperparameter. The alignment term minimizes the angular distance between genuine audio-ultrasound pairs, while the hinge-based misalignment term forces misaligned pairs to have similarity below $\delta$. 
    This asymmetric design prevents collapse and enforces speech–ultrasound synchronization.

    \item \textbf{Perceptual Loss:} To ensure imperceptibility, we employ a combination of speech quality losses. The total perceptual loss \(\mathcal{L}_{\text{percept}}\) is a weighted sum of complementary spectral and energy-based objectives:
\begin{equation}
\begin{aligned}
\mathcal{L}_{\text{percept}} = \,
& \lambda_{\text{mel}} \mathcal{L}_{\text{mel}} + \lambda_{\text{msspec}} \mathcal{L}_{\text{msspec}} \\
& + \lambda_{\text{sisnr}} \mathcal{L}_{\text{sisnr}} + \lambda_{\text{loud}} \mathcal{L}_{\text{loud}}.
\end{aligned}
\end{equation}

Specifically, we incorporate: (i) a mel-spectrogram loss \(\mathcal{L}_{\text{mel}}\) that operates on perceptually-scaled frequency representations; (ii) a multi-resolution STFT loss  \(\mathcal{L}_{\text{msspec}}\) that captures spectral fidelity across multiple time-frequency resolutions; (iii) a scale-invariant signal-to-noise ratio (SI-SNR) loss \(\mathcal{L}_{\text{sisnr}}\) that measures energy-normalized distortion independent of signal amplitude; and (iv) a time-frequency loudness ratio loss \(\mathcal{L}_{\text{loud}}\) that penalizes perceptually salient energy deviations across frequency bands between the original and watermarked audio. This multi-faceted objective ensures that the embedded watermark remains inaudible while preserving the acoustic characteristics of the original speech.

\end{itemize}


\noindent\textbf{End-to-end Training and Detection.}
During training, the generator receives the raw waveform $x$ and produces both a watermark signal $w$ and an extracted speech $x_a$. The watermarked audio $x_w$ is constructed as:
$
x_w = x_a + \alpha w$,
where $\alpha$ is a scalar controlling watermark strength. Gradients from the localization, alignment, and perceptual losses are backpropagated through the generator--detector pipeline. We monitor gradient norms and apply gradient clipping with a balancer mechanism to prevent any objective from dominating training. Together, these techniques enable stable and efficient training optimized for accuracy, robustness, and imperceptibility under real-world acoustic and adversarial conditions.

 \begin{figure}
     \centering
     \includegraphics[width=0.65\linewidth]{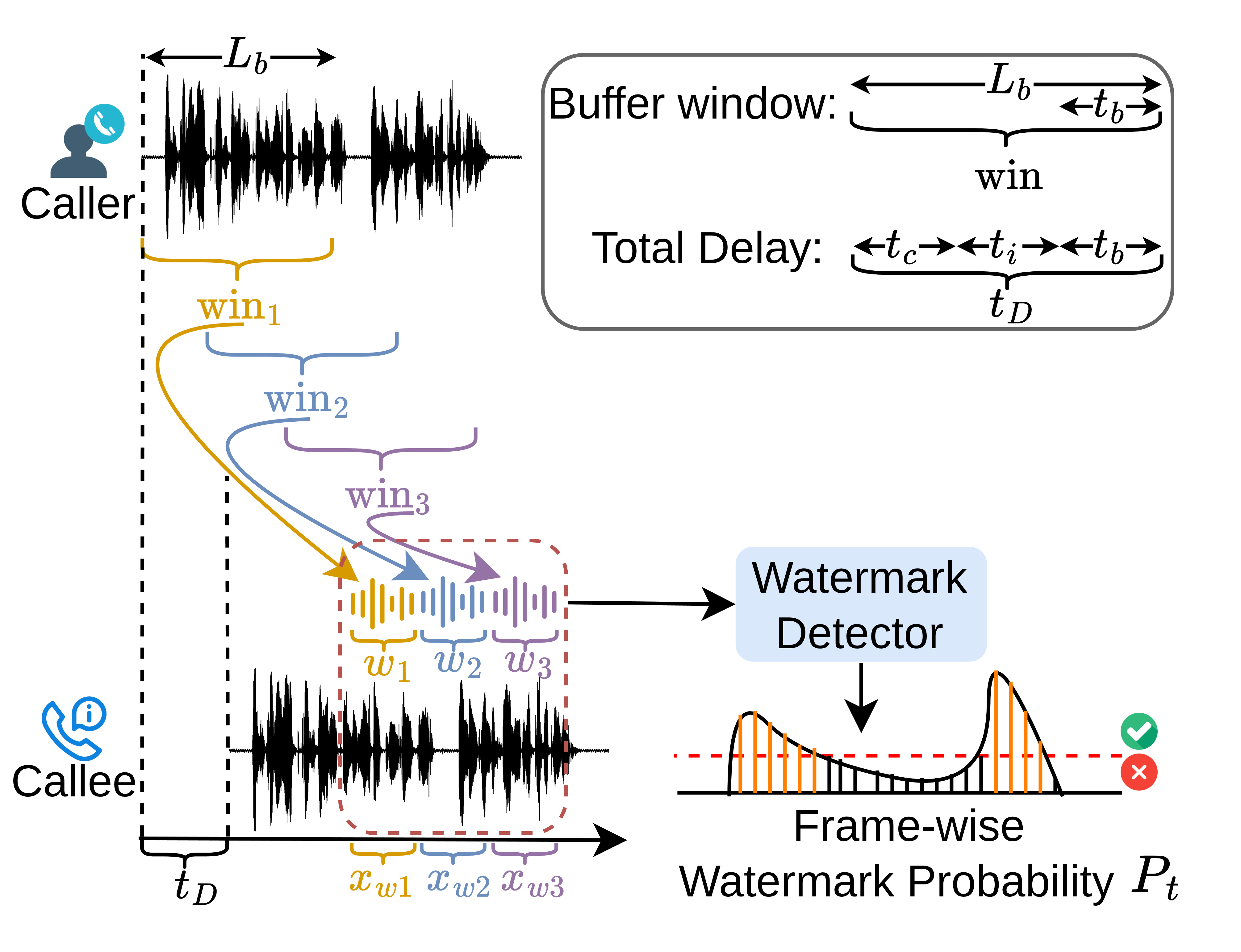}
     \caption{Sliding-window watermark generation and detection for real-time calls.}
     \label{fig:real_time}
 \end{figure}  
\subsection{Real-time Implementation}
We design a real-time streaming architecture that enables the generator and detector to operate continuously during live phone calls under strict latency and compute constraints. Unlike offline watermarking, in-call watermarking must process speech as it arrives while minimizing buffering delays and remaining robust to natural variations. Our solution adopts a sliding-window pipeline at both endpoints, coordinating buffering, window stepping, and Root Mean Square (RMS) energy--based speech gating to reduce computation while preserving seamless audio flow. As shown in Fig.~\ref{fig:real_time}, the generator processes the audio stream in fixed increments to produce short watermarked segments, while the detector aligns its sliding window with the generator's delay budget to provide low-latency, frame-wise confidence estimates. 

\setlength{\textfloatsep}{17pt}
\begin{algorithm}[!t]
\footnotesize
\caption{Real-time Watermark Generation}
\label{alg:sliding_window}
\KwIn{Audible signal stream $S(t)$, buffer length $L_b$, step size $t_b$}
\KwOut{Continuous watermarked audio stream $W(t)$}

Initialize an empty buffer $B \leftarrow []$\;
\While{the call is active}{
    Append incoming samples from $S(t)$ to $B$\;
    \If{$|B| \ge L_b$}{
        $A_{\mathrm{wm}} \leftarrow \mathrm{Generator}(B)$\;
        $A_{\mathrm{send}} \leftarrow A_{\mathrm{wm}}[L_b - t_b : L_b]$\;
        
        Transmit $A_{send}$ to callee device\;
        
        Discard first $t_b$ samples in $B$\;
    }
}
\end{algorithm}

\noindent\textbf{Sliding-window Watermark Generation.}
As shown in Algorithm~\ref{alg:sliding_window}, incoming speech samples are accumulated in a buffer of length $L_b$ on the caller device. During an initial warm-up phase, speech is transmitted without watermarking until the first buffer is filled, providing sufficient context for watermark generation. The full buffer is then fed into the generator, but only the last $t_b$ samples, corresponding to the newest watermarked speech, are transmitted, where $t_b$ denotes the step size. The window then advances by $t_b$, and each subsequent $L_b$-sample window generates a new $t_b$-sample watermarked segment.

\noindent\textbf{RMS-based Speech Gating.}
We apply an RMS threshold of 0.05 to detect speech and reduce unnecessary computation. Segments with RMS below 0.05 are treated as silence and passed without watermarking. This improves efficiency, prevents bias from silence, and maintains predictable latency.

\noindent\textbf{Window-aligned Watermark Detection.} We apply the same sliding-window algorithm during watermark detection. Once the detector buffer reaches length $L_b$, it is processed to produce a per-frame confidence map. Because the earliest portion corresponds to audio transmitted before watermarking begins, these frames are excluded from confidence computation. Let
$
n = \left\lfloor L_b/t_b \right\rfloor
$
denote the number of sliding steps before the buffer becomes fully watermarked. During the first $n$ steps of a new call session, only the newest $t_b$ watermarked region contributes to the confidence map, while the initial unwatermarked segment of length
$
L_b - n \times t_b
$
is excluded. After $n$ steps, confidence is computed over the entire buffer. We apply the same RMS threshold during detection, excluding silent frames from detection.

\noindent\textbf{Confidence Estimation.}
For each newly filled window, the detector produces an \emph{instant confidence level} reflecting the model's output. Because this value may fluctuate due to transient noise or distortions, we compute an \emph{averaged confidence level} over the most recent $N$ instant values after removing the largest and smallest values. This stabilization suppresses outliers and reduces false alarms while preserving sustained confidence drops indicative of attacks. The detector app displays both confidence levels in real time and tracks the lowest averaged confidence over the call to indicate manipulation.

\noindent\textbf{Latency Modeling and Synchronization.}
To compensate for watermark generation and transmission delays, the callee delays playback by
$
t_D = t_b + t_i + t_c,
$
where $t_i$ is generator inference time and $t_c$ is call transmission latency. After this delay, detection proceeds with the same window size $L_b$ and step size $t_b$, maintaining synchronization with the generator.

\subsection{Cross-platform Deployment}

We deploy our system end-to-end on commodity mobile devices via native iOS and Android applications. The generator and detector models are exported to platform-compatible formats, enabling on-device inference without external servers. Communication between devices is established through IP-based calling, which streams the watermarked speech from the caller device to the callee device.

\noindent\textbf{Cross-Platform On-Device Inference.}
To ensure compatibility, the generator and detector are converted to lightweight formats. The Android app uses NCNN~\cite{Ni_ncnn_2017} for model acceleration, while the iOS app uses Core ML optimized for real-time audio processing. Both applications perform inference on-device, avoiding cloud services and preserving privacy. Since standard signal processing libraries such as \texttt{torchaudio} are unavailable or inefficient on mobile platforms, we implement mobile-compatible alternatives for FIR filtering, linear resampling, and model export, enabling native execution on both Android and iOS without additional hardware. We will open-source these mobile implementations upon publication to facilitate reproducibility.

    \begin{table}
    \centering
    \footnotesize
    \begin{tabular}{l p{1.8cm} >{\centering\arraybackslash}p{2.8cm}}
        \hline
        \textbf{Phone Model}& \textbf{OS}& \textbf{Ultrasound Gain (dB)}\\
        \hline
        Motorola edge 5G UW 1 & Android 13 & 0.47 \\
        Motorola edge 5G UW 2 & Android 13 & 0.39 \\
        Google pixel 9 & Android 16 & 3.64 \\
        Sony Xperia XA1 Plus & Android 7 & 3.72 \\
        iPhone 17 Pro & iOS 26.2.1 & 5.78 \\
        iPhone 16 Pro Max & iOS 18.6.2 & 4.94 \\
        iPhone 14 Pro Max & iOS 26.2.1 & 5.27 \\
        iPhone 13 Mini & iOS 26.2.1 & 4.20\\
        \hline
    \end{tabular}
    \caption{Smartphones used for data collection.}
    \label{tab:phone_ultrasound_gain}
\end{table}

\section{Evaluation}\label{sec-evaluation}
\subsection{Experiment Setup}

\noindent\textbf{Datasets.}
Our dataset contains over 2,000 utterances, including over 1,700 genuine recordings from 22 speakers, 150 TTS samples, and 150 VC samples. We partition the data into training, validation, and evaluation sets with an 80\%/10\%/10\% ratio. All spoofing instances are constructed independently within each split to prevent data leakage. Positive samples are generated from genuine recordings and include both clean and in-call-distorted watermarked speech. All signals are segmented into two-second samples at 48~kHz.

\noindent\textbf{Genuine Dataset.}
We develop custom Android and iOS applications that simultaneously emit ultrasonic sensing signals and record audible speech and ultrasonic reflections, enabling synchronized multimodal capture without external hardware. During data collection, participants read a fixed set of ten sentences representative of common phone-scam scenarios, including urgent account warnings, payment requests, and legal threats. We collect over 1,700 genuine recordings from 22 participants using four iPhones and four Android smartphones across four environments (home, office, laboratory, and outdoor) and two call modes (speaker and earpiece). As summarized in Table~\ref{tab:phone_ultrasound_gain}, the devices span different operating systems and ultrasound gains from 0.39~dB to 5.78~dB. Data across devices, environments, and call modes are combined during training to improve robustness to device heterogeneity and usage variations.

\noindent\textbf{Synthesized Dataset.}
We use the same speech scripts for genuine, TTS, and VC samples to ensure consistent linguistic content across attack types. For each speaker, we select and resample one representative genuine recording as the speaker reference for TTS voice cloning and VC. We generate 150 TTS samples using WaveNet~\cite{van2016wavenet}, Tacotron~2~\cite{shen2018natural}, and FastSpeech~\cite{ren2019fastspeech}, and 150 VC samples using VAE-based~\cite{long2022enhancingzeroshotvoiceconversion}, GAN-based~\cite{li2021starganv2vcdiverseunsupervisednonparallel}, and diffusion-based models~\cite{choi2023diffhiervcdiffusionbasedhierarchicalvoice}. Speech scripts are detailed in Appendix~\ref{app:data_generation}.

\noindent\textbf{Spoof Dataset.}
After partitioning the data, we construct spoofing instances independently within each split. The spoof speech pool consists of over 1,700 replay utterances, 150 TTS utterances, and 150 VC utterances. These speech samples are combined with ultrasound according to the attack scenarios described below, ensuring that no cross-split speech--ultrasound pairing occurs.

\noindent\textbf{Attack Setup.}
Each spoofed speech consists of audible speech and ultrasound signals. \textit{Scenario 1: Audio-only Spoofing.} Spoofed speech is provided without ultrasound. \textit{Scenario 2: Random Ultrasound Spoofing.} Spoofed speech is paired with an unrelated ultrasound waveform. \textit{Scenario 3: Speech-synchronized Ultrasound Spoofing.} Spoofed speech is paired with ultrasound from a different speaker uttering the same sentence. RMS-based segmentation aligns utterance onset, producing temporally synchronized but physically mismatched speech--ultrasound pairs. \textit{Scenario 4: Adaptive Adversaries.} Adaptive attacks are evaluated separately after training by perturbing existing detector inputs through model inference or optimization and therefore require no additional spoofing corpus.

\noindent\textbf{In-call Distortions.}
To improve and evaluate robustness under realistic phone-call conditions, we apply in-call distortions during both training and evaluation. For each watermarked speech sample, we randomly select one distortion before detection, including low-pass filtering (5~kHz cutoff), high-pass filtering (500~Hz cutoff), band-pass filtering (300~Hz--8~kHz), up-down resampling via 32~kHz, MP3 and AAC compression at 128~kbps, pink ($\sigma=0.01$) and Gaussian ($\sigma=0.001$) noise, $\pm20$~dB gain changes, moving-average smoothing with 2--10 sample windows, echo with volume 0.1--0.5 and delay 0.1--0.5~s, and Encodec compression with 16 quantization levels followed by resampling.

\noindent\textbf{Training Setup.}
We train the model on an NVIDIA RTX 5090 GPU with 32~GB memory using the AudioCraft framework~\cite{copet2023simple}, adapted for our data loading and training pipeline. We use Adam with $\beta=(0.5,0.9)$, an initial learning rate of $5\times10^{-4}$ with adaptive scheduling, gradient clipping at a maximum norm of 1.0, and exponential moving average with a decay of 0.99. Training uses FP32 precision, a batch size of 16, and 700k steps. Loss weights are set to $\lambda_{\text{loc}}=10.0$, $\lambda_{\text{align}}=10.0$, $\lambda_{\text{mel}}=1.0$, $\lambda_{\text{msspec}}=2.0$, $\lambda_{\text{sisnr}}=1.0$, and $\lambda_{\text{loud}}=10.0$ to balance detection accuracy and perceptual quality. After training, the generator and detector are deployed on commodity smartphones for on-device inference at the caller and callee side, respectively.

\begin{figure}[t]
\centering

\begin{subfigure}[t]{0.48\columnwidth}
\centering
\includegraphics[width=\linewidth]{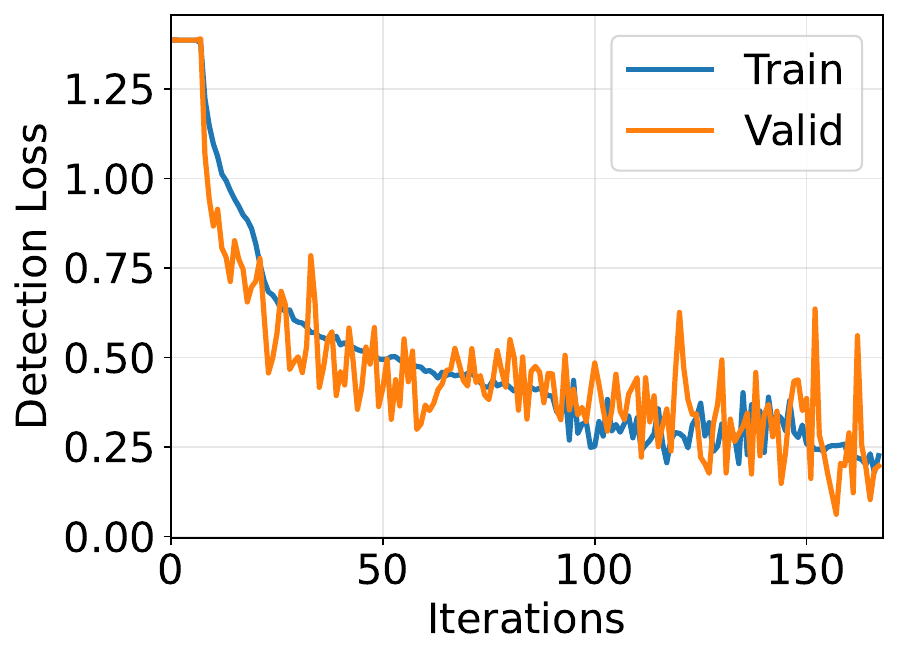}
\caption{Detection loss.}
\label{fig:detection_loss}
\end{subfigure}
\hfill
\begin{subfigure}[t]{0.48\columnwidth}
\centering
\includegraphics[width=\linewidth]{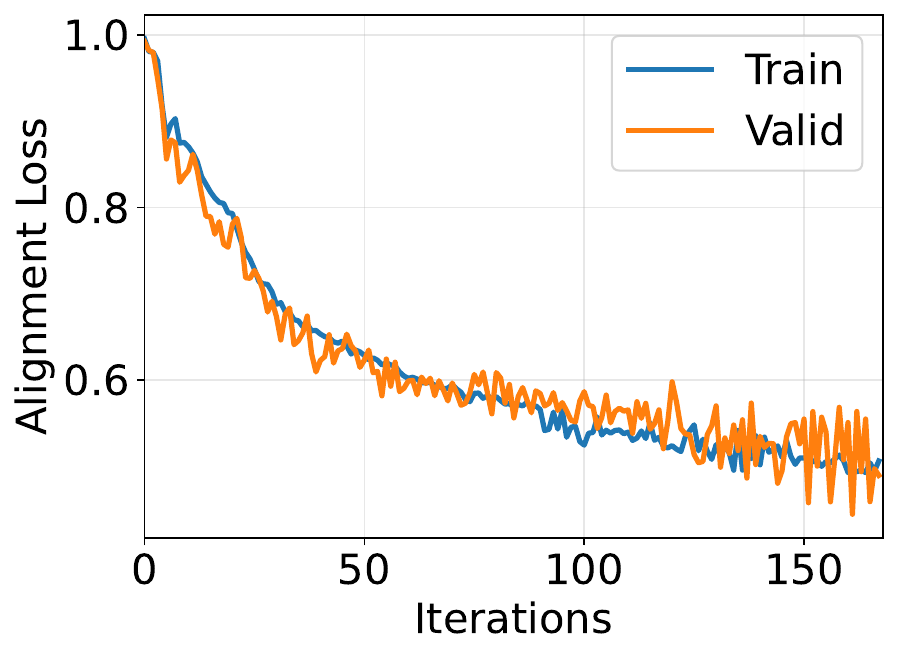}
\caption{Alignment loss.}
\label{fig:alignment_loss}
\end{subfigure}

\caption{Training loss over iterations.}
\label{fig:detection_alignment_loss}
\end{figure}

\noindent\textbf{Evaluation Metrics.}
We evaluate \ours in terms of detection performance and speech quality. For detection, we report Accuracy (Acc), True Positive Rate (TPR), False Positive Rate (FPR), Attack Success Rate (ASR), and the Receiver Operating Characteristic (ROC) curve with Area Under the Curve (AUC). Accuracy measures overall classification correctness, TPR measures the fraction of watermarked samples correctly detected, FPR measures the fraction of non-watermarked samples incorrectly classified as watermarked, and ASR measures the fraction of attack attempts successfully accepted as legitimate. The ROC curve characterizes the TPR--FPR tradeoff across detection thresholds, while AUC summarizes overall detection performance. For speech quality, we report Perceptual Evaluation of Speech Quality (PESQ) for perceptual quality, Short-Time Objective Intelligibility (STOI) for speech intelligibility, and Scale-Invariant Signal-to-Noise Ratio (SI-SNR) for reconstruction quality, with higher values indicating better performance.

\begin{figure}[t]
\centering
\begin{subfigure}[b]{0.48\columnwidth}
    \centering
    \includegraphics[width=\linewidth]{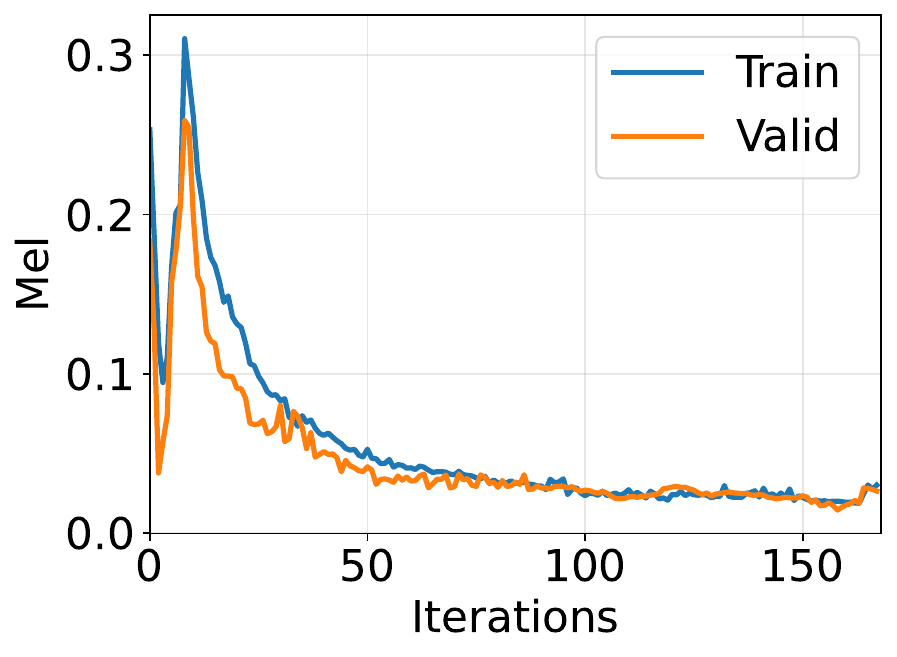}
    \caption{Mel-spectrogram.}
    \label{fig:quality_mel}
\end{subfigure}
\hfill
\begin{subfigure}[b]{0.48\columnwidth}
    \centering
    \includegraphics[width=\linewidth]{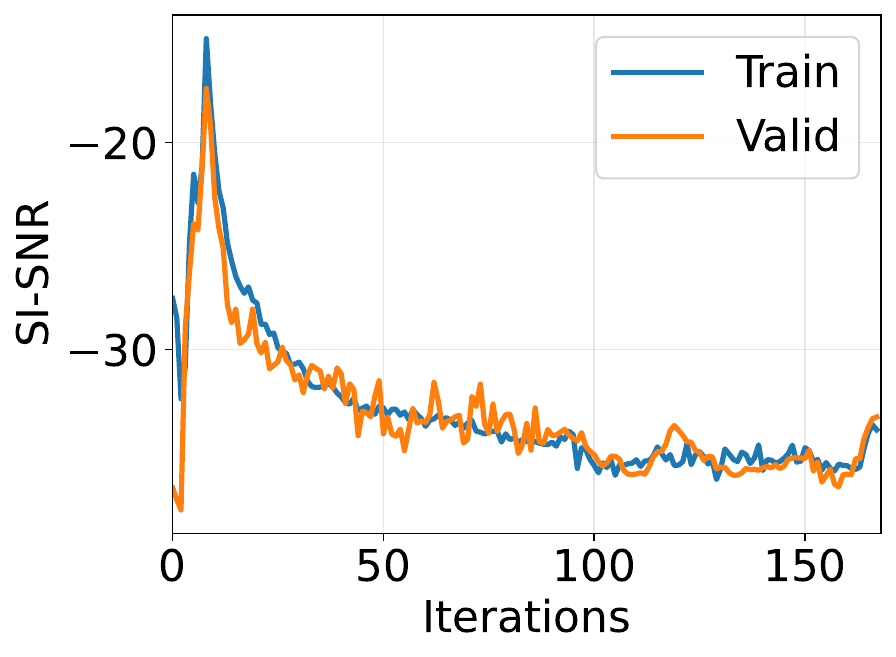}
    \caption{SI-SNR.}
    \label{fig:quality_sisnr}
\end{subfigure}

\begin{subfigure}[b]{0.48\columnwidth}
    \centering
    \includegraphics[width=\linewidth]{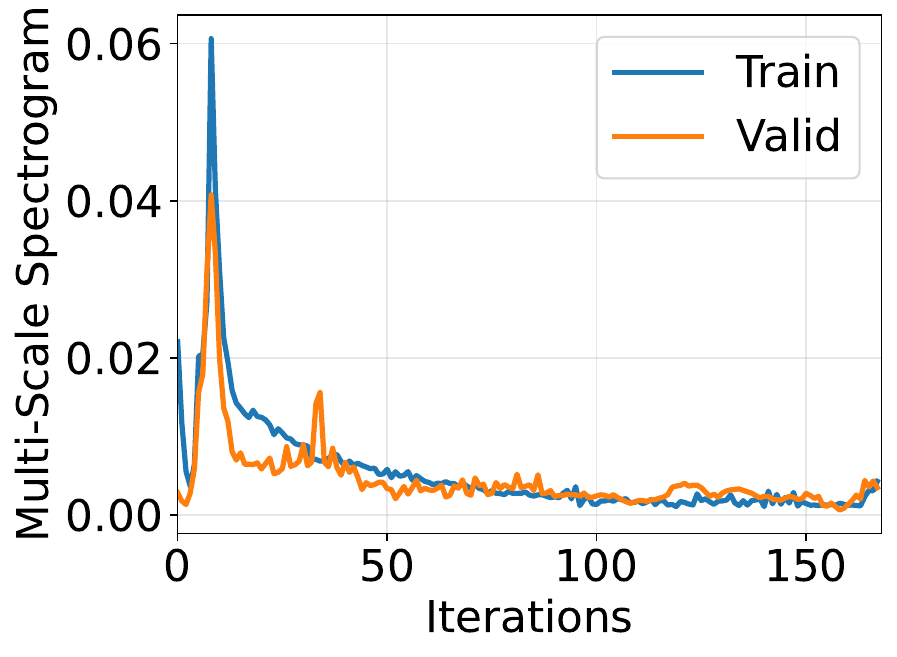}
    \caption{Multi-Scale Spectrogram.}
    \label{fig:quality_msspec}
\end{subfigure}
\hfill
\begin{subfigure}[b]{0.48\columnwidth}
    \centering
    \includegraphics[width=\linewidth]{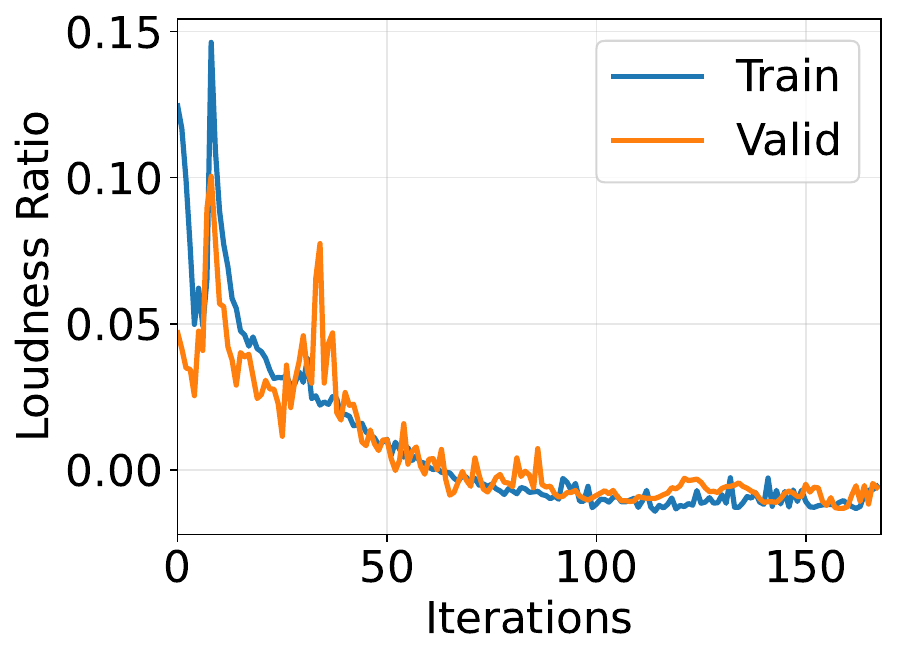}
    \caption{Loudness Ratio.}
    \label{fig:quality_loudness}
\end{subfigure}

\caption{Quality losses over iterations.}
\label{fig:quality_metrics}

\end{figure}

\subsection{Model Training Performance}
We jointly train the watermark generator and detector in an end-to-end manner. Positive samples are watermarked audio generated from aligned speech--ultrasound pairs, while negative samples include unwatermarked audio and watermarked audio generated from misaligned pairs. The training objective jointly optimizes watermark detectability, attack robustness, and perceptual audio quality.

\noindent\textbf{Loss Convergence.}
Fig.~\ref{fig:detection_alignment_loss} shows steadily decreasing detection and alignment losses with small train--validation gaps, indicating stable optimization and good generalization.

\noindent\textbf{Audio Quality.}
Fig.~\ref{fig:quality_metrics} shows that the quality-related losses (Mel-spectrogram, SI-SNR, MSSpec, and loudness) converge steadily. Their initial increases occur as detection and alignment objectives dominate, introducing larger perturbations; as detectability improves, the quality objectives progressively reduce perceptual distortion. The small train--validation gap further indicates good generalization.

\begin{figure*}[htbp]
\centering

\begin{minipage}[t]{0.2479\textwidth}
  \centering

  \includegraphics[width=0.97\linewidth]{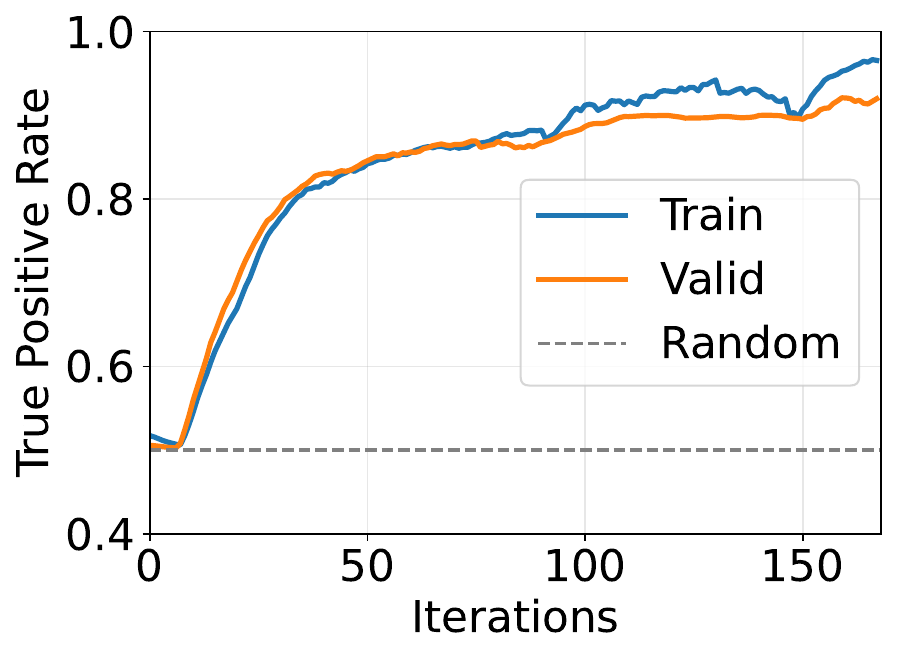}
  \includegraphics[width=0.97\linewidth]{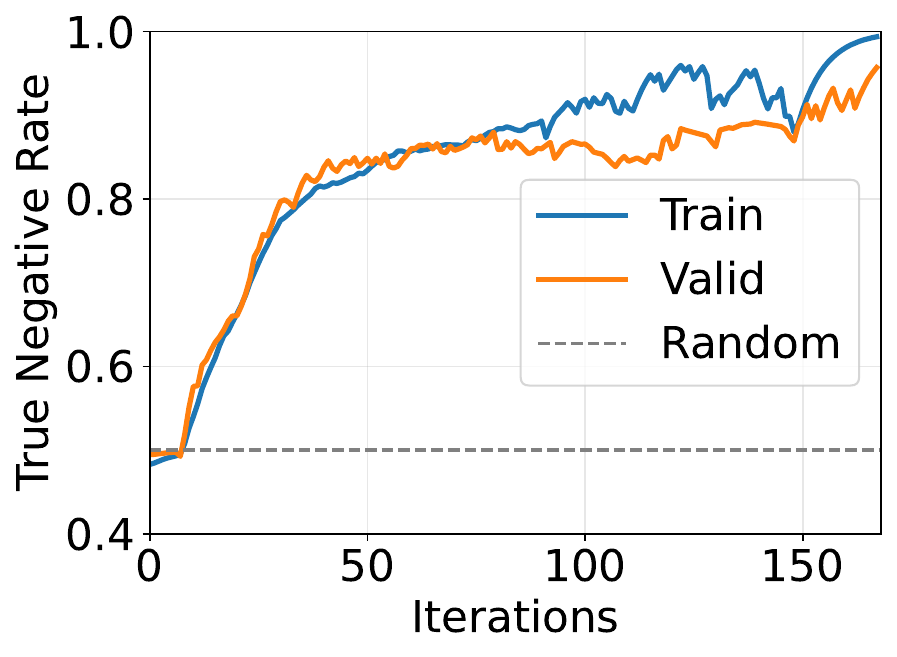}

  \caption{Training TPR/TNR.} 
  \label{fig:col1}
\end{minipage}
\begin{minipage}[t]{0.746\textwidth}
  \centering
  
  \includegraphics[width=0.32\linewidth]{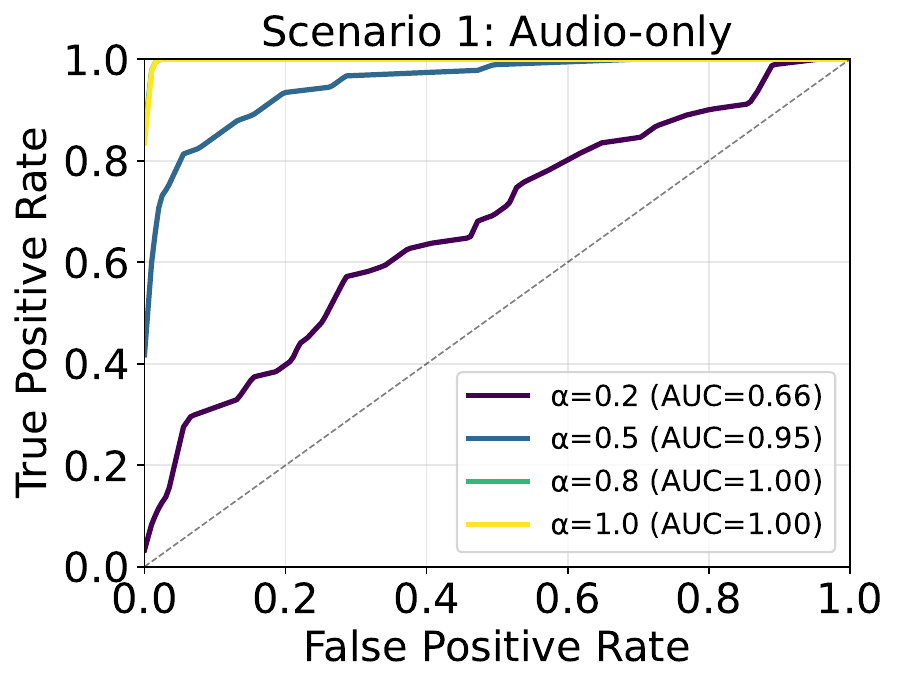}\hfill
  \includegraphics[width=0.32\linewidth]{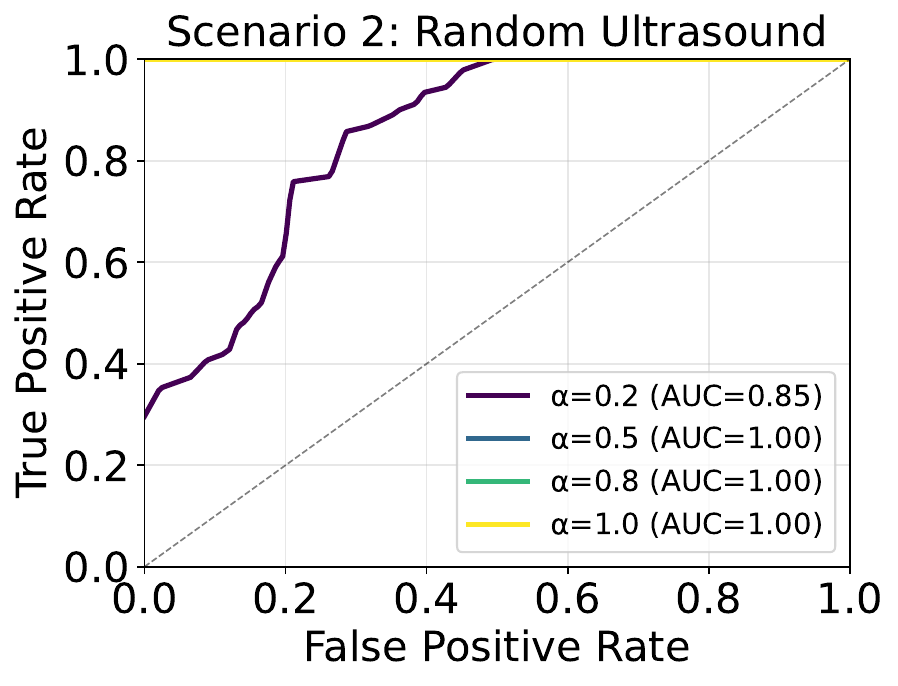}\hfill
  \includegraphics[width=0.32\linewidth]{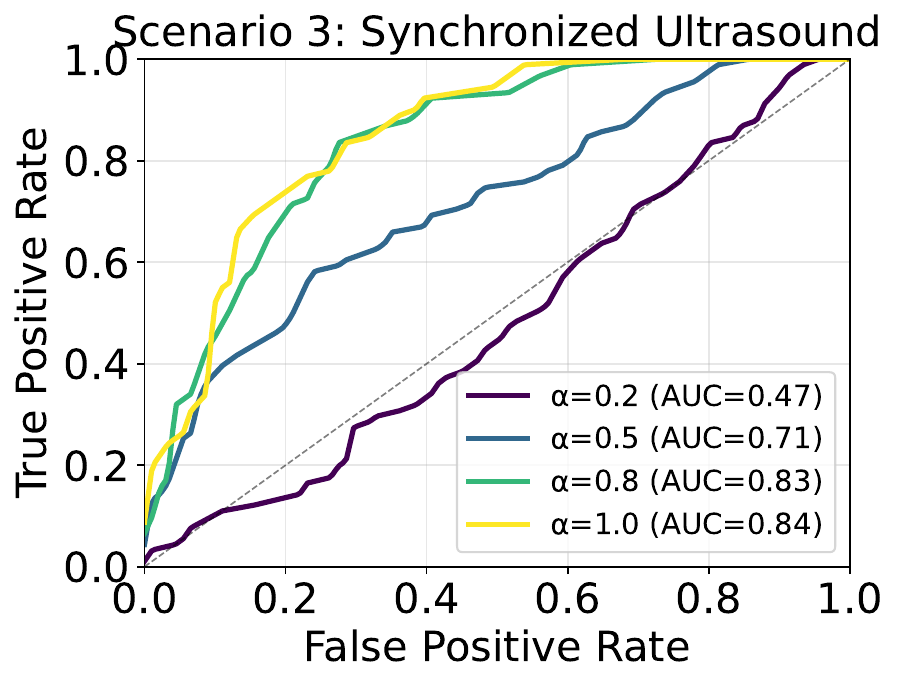}

  \includegraphics[width=0.32\linewidth]{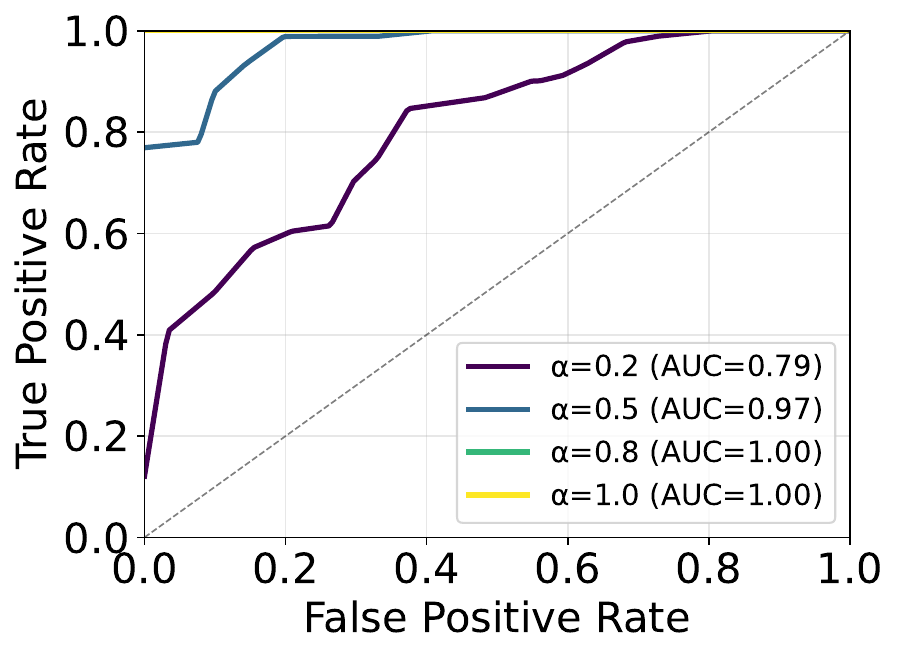}\hfill
  \includegraphics[width=0.32\linewidth]{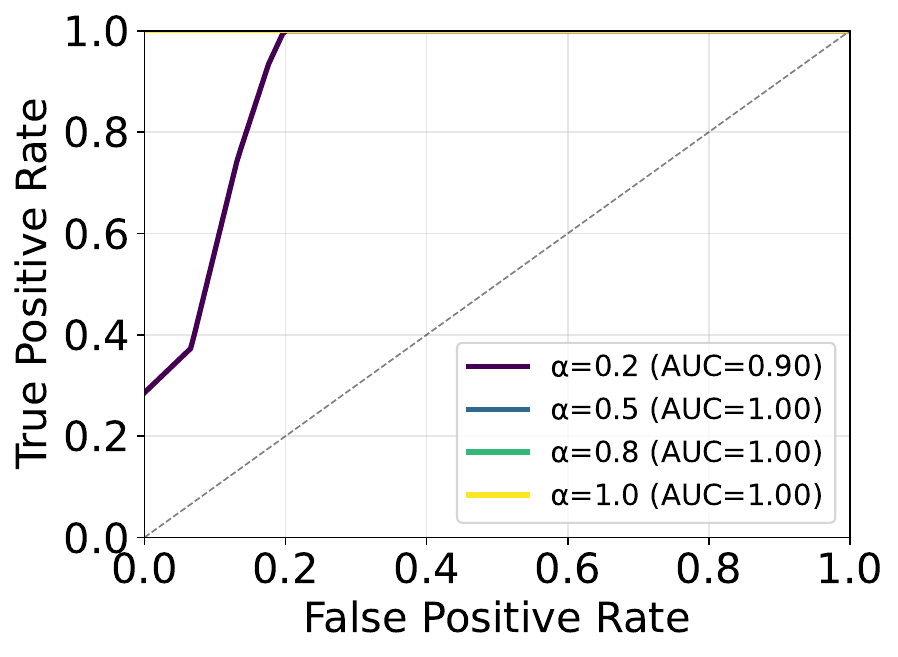}\hfill
  \includegraphics[width=0.32\linewidth]{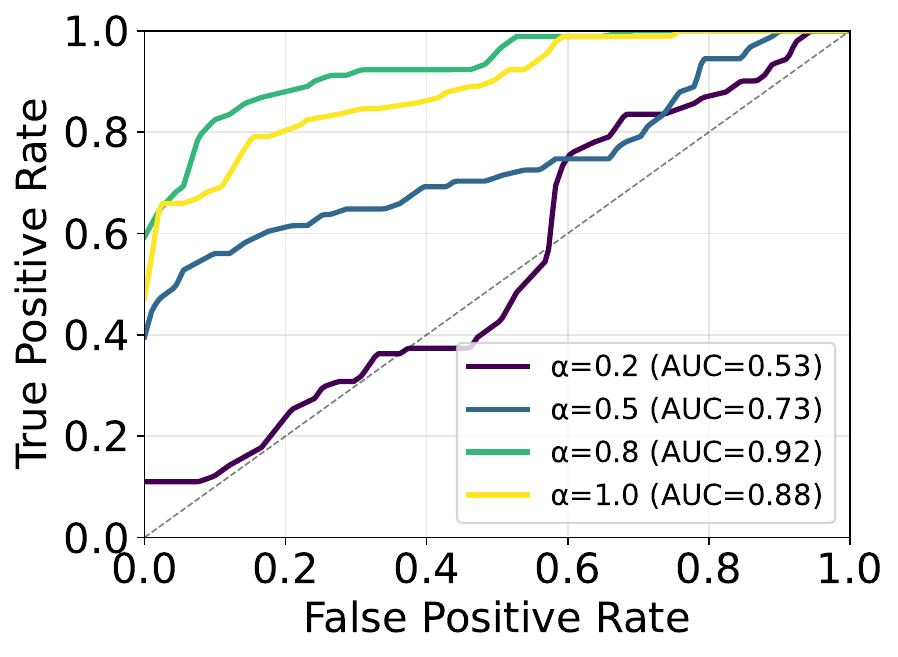}
  \caption{ROC curves for replay (top) and synthesis (bottom) attacks.}
   \label{fig:col234}
   
\end{minipage}

\label{fig:training_and_roc}
\end{figure*}

\noindent\textbf{Watermark Detection Performance.}
Fig.~\ref{fig:col1} shows that TPR and TNR improve steadily during training, reaching 0.93 and 0.99, respectively. Both remain well above the 0.5 random baseline, demonstrating effective discrimination between positive and negative samples.

\begin{table}[t]
\centering
\footnotesize
\begin{tabular}{lcc}
\toprule
\textbf{Attack Type} & \textbf{AUC} & \textbf{FPR} \\
\midrule
Audio-only + replay                 & 1.00 & 0.01 \\
Audio-only + synthesis              & 1.00 & 0.01 \\
Random ultrasound + replay          & 1.00 & 0.00 \\
Random ultrasound + synthesis       & 1.00 & 0.00 \\
Synchronized ultrasound + replay    & \textbf{0.84} & \textbf{0.37} \\
Synchronized ultrasound + synthesis & \textbf{0.88} & \textbf{0.30} \\
\midrule
\textbf{Average}                    & \textbf{0.95} & \textbf{0.12} \\
\bottomrule
\end{tabular}
\caption{Detection results against attack types ($\alpha=1.0$).}
\label{tab:attack_results}
\end{table}

\subsection{Performance Under Spoofing Attacks}\label{sec:pusa}
We evaluate \ours against three spoofing scenarios. For each scenario, the audible input is either replayed or synthesized using TTS or VC, resulting in six attack types. Table~\ref{tab:attack_results} summarizes the detection performance, while Fig.~\ref{fig:col234} show the impact of watermark strength $\alpha$.

\begin{itemize}[leftmargin=*]
\item \textbf{Audio-only Attacks:}
For replay and synthesis attacks, \ours achieves near-perfect detection, with an AUC of 1.00 and FPR of 0.01 in both cases. Fig.~\ref{fig:col234} shows that increasing $\alpha$ improves separability, with $\alpha \geq 0.8$ achieving near-perfect detection.

\item \textbf{Random Ultrasound Attacks:}
When unrelated ultrasound is combined with spoofed audible speech, we achieve an AUC of 1.00 with zero false positives for both replay and synthesis. As shown in Fig.~\ref{fig:col234}, detection remains robust even at $\alpha=0.2$, indicating that unrelated ultrasonic motion does not produce valid watermark signatures.

\item \textbf{Synchronized Ultrasound Attacks:}
Synchronized ultrasound poses the strongest challenge, reducing AUC to 0.84 for replay and 0.88 for synthesis, with FPRs of 0.37 and 0.30, respectively. Fig.~\ref{fig:col234} show that detection improves substantially as $\alpha$ increases: performance approaches chance at $\alpha=0.2$ but becomes substantially more separable at $\alpha=0.8$ and $1.0$ with limited audio-quality degradation. These results suggest that residual inconsistencies between spoofed speech and mimicked articulatory motion remain detectable at sufficient watermark strength.
\end{itemize}

\subsection{Performance Under Adaptive Attacks}
We further evaluate \ours against \emph{adaptive adversaries} that optimize inputs to evade watermark detection. Unlike spoofing attacks, these adversaries exploit knowledge of the detection objective to reduce watermark confidence through structured perturbations.

\noindent\textbf{Adaptive Adversary Model.}
We consider three adversaries with increasing capabilities: (i) \textit{Black-box}, with query access but no knowledge of model parameters or gradients; (ii) \textit{Grey-box}, with knowledge of the architecture, training pipeline, and loss functions but not the trained weights; and (iii) \textit{White-box}, with full access to the model and gradients. We evaluate five attacks spanning these capabilities: Gaussian noise, query-distilled transfer, live VC with genuine articulation, recipe-aware transfer, and white-box gradient optimization. Table~\ref{tab:rev:attack_property_matrix} summarizes their capabilities and real-time feasibility. We use a detection threshold of 0.5 and define ASR as the fraction of spoofed inputs accepted as legitimate.

\begin{table}[t]
\centering
\scriptsize
\setlength{\tabcolsep}{2.5pt}
\begin{tabular}{lccccc}
\toprule
\textbf{Attack} & \textbf{Capability} & \textbf{Detector} & \textbf{Detector} & \textbf{Detector} & \textbf{Real-Time} \\
\textbf{Method} & \textbf{Level} & \textbf{Queries} & \textbf{Arch.} & \textbf{Weights} & \textbf{Feasible} \\
\midrule
Gaussian noise     & \textbf{Baseline}  & \xmark & \xmark & \xmark & \cmark \\
Query transfer     & \textbf{Black-box} & \cmark & \xmark & \xmark & \xmark \\
Live VC            & \textbf{Black-box} & \xmark & \xmark & \xmark & \cmark \\
Recipe transfer    & \textbf{Grey-box}  & --     & \cmark & \xmark & \xmark \\
APGD               & \textbf{White-box} & --     & \cmark & \cmark & \xmark \\
\bottomrule
\end{tabular}
\caption{Attacker capabilities for adaptive attacks.}
\label{tab:rev:attack_property_matrix}
\end{table}

\begin{figure}[t]
\centering
\includegraphics[width=0.8\linewidth]{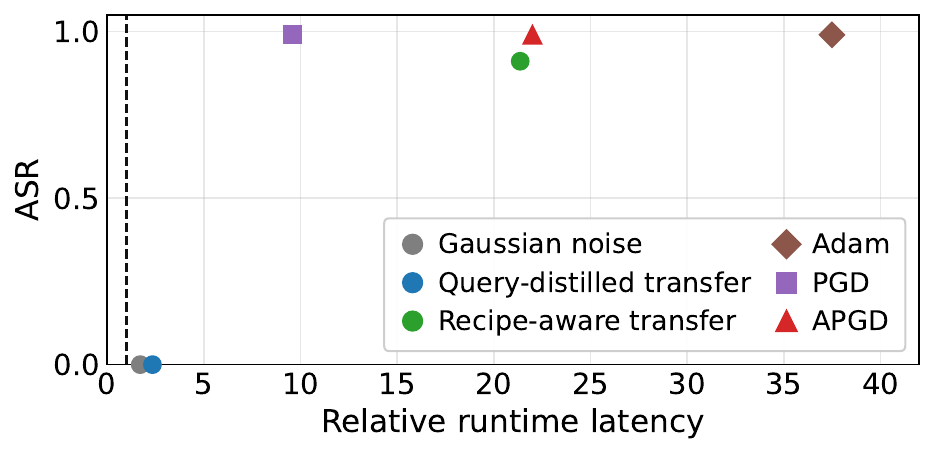}
\caption{Real-time feasibility of adversarial attacks. Runtime normalized to end-to-end inference time.}
\label{fig:rev:realtime_efficiency}
\end{figure}

\begin{itemize}[leftmargin=*]
\item \textbf{Gaussian Noise:}
As a non-adaptive baseline, we add zero-mean white Gaussian noise with standard deviation $\sigma \in [5\times10^{-4},\,2\times10^{-1}]$. ASR remains 0 across all noise levels, indicating that unstructured perturbations do not induce false acceptance.

\item \textbf{Query-distilled Transfer:}
For black-box attacks, the attacker uses limited detector queries ($Q\in\{100,200,300,500\}$) to train a surrogate via score matching and applies Projected Gradient Descent (PGD) to generate transferable perturbations. ASR remains below 0.05 across all query budgets. Even at $Q=500$, the Pearson correlation between surrogate and target outputs remains below 0.16, indicating poor approximation of the target decision boundary and limited transferability.

\item \textbf{Live VC with Genuine Articulation:}
The attacker uses live VC to transform their speech into a target voice while retaining live ultrasonic articulation. Table~\ref{tab:rev:vcpipe} shows that the fastest model (YourTTS\&VC~\cite{casanova2022yourtts}) incurs $3.6\times$ latency overhead comparing to watermark detector inference latency, while FreeVC~\cite{li2023freevc} and Seed-VC~\cite{li2022freevchighqualitytextfreeoneshot} incur $6.9\times$ and $13.4\times$, respectively. It incurs substantial overhead and speech--ultrasound misalignment during live calls.

\item \textbf{Recipe-aware Transfer:}
For grey-box attacks, the attacker trains an identical surrogate with different weights and applies PGD to generate transferable perturbations. As shown in Table~\ref{tab:rev:merged_attack}, transfer ASR remains low at small perturbation budgets (0.06 and 0.16 for $\varepsilon \leq 2\times10^{-3}$), but reaches 0.81 at $\varepsilon=10^{-2}$ and exceeds 0.85 at larger budgets. Thus, recipe-level knowledge improves transferability primarily under stronger, potentially perceptible perturbations.

\item \textbf{White-box Attacks:}
With full model access, the attacker directly optimizes bounded waveform perturbations $\delta\in[-\varepsilon,\varepsilon]^T$. To rule out gradient masking, we evaluate Adam, PGD~\cite{madry2019deeplearningmodelsresistant}, and Auto-PGD (APGD)~\cite{croce2020reliableevaluationadversarialrobustness}. Table~\ref{tab:rev:merged_attack} shows consistent ASRs across optimizers, differing by at most 0.04 at $\varepsilon=10^{-3}$ and 0.01 for $\varepsilon\geq2\times10^{-3}$. ASR remains relatively low ($\approx0.3$) at small budgets ($\varepsilon\leq10^{-3}$) but approaches 1.0 as $\varepsilon$ increases, showing that sufficiently strong white-box perturbations can bypass the detector.

\item \textbf{Real-Time Feasibility:}
Fig.~\ref{fig:rev:realtime_efficiency} compares attack success and runtime normalized to our system's end-to-end inference time. Gaussian noise and query-distilled transfer are computationally inexpensive but achieve low ASR, whereas effective recipe-aware and white-box attacks incur $>10\times$ runtime overhead. This tradeoff limits effective adaptive attacks under live-call latency constraints.

\end{itemize}

\begin{table}[t]
\centering\small
\setlength{\tabcolsep}{4pt}
\begin{tabular}{lccc}
\toprule
\textbf{Method} & \textbf{Synthesis (ms)} & \textbf{Detector (ms)} & \textbf{Overhead} \\
\midrule
YourTTS\&VC & 36.1  & 15.3 & $\mathbf{3.6\times}$ \\
FreeVC      & 85.3  & 15.3 & $\mathbf{6.9\times}$ \\
Seed-VC     & 178.6 & 15.3 & $\mathbf{13.4\times}$ \\
\bottomrule
\end{tabular}
\caption{Runtime overhead of live VC attacks.}
\label{tab:rev:vcpipe}
\end{table}

\begin{table}[t]
\centering
\footnotesize
\begin{tabular}{lccccc}
\toprule
& \multicolumn{3}{c}{\textbf{White-box ASR}} & \multicolumn{2}{c}{\textbf{Grey-box ASR}} \\
\textbf{$\varepsilon$} & Adam & PGD & APGD & Surrogate & Transfer \\
\midrule
5e-4 & 0.07 & 0.07 & 0.07 & -- & -- \\
1e-3 & \textbf{0.27} & \textbf{0.31} & \textbf{0.31} & 0.08 & \textbf{0.06} \\
2e-3 & 0.75 & 0.75 & 0.76 & 0.35 & \textbf{0.16} \\
5e-3 & 0.97 & 0.98 & 0.98 & 0.88 & 0.61 \\
1e-2 & 0.99 & 0.99 & 0.99 & 1.00 & \textbf{0.81} \\
2e-2 & 0.99 & 0.98 & 0.99 & 0.99 & 0.86 \\
5e-2 & 1.00 & 1.00 & 1.00 & 0.92 & 0.91 \\
\bottomrule
\end{tabular}
\caption{ASR under different perturbation budgets $\varepsilon$ for white-box and grey-box attacks.}
\label{tab:rev:merged_attack}
\end{table}

\subsection{Comparison with Liveness Detection}
We compare our system with recent \textbf{audio-only liveness detection} methods. Prior work detect spoofing artifacts from audible speech~\cite{wu15e_interspeech,Zhang_2023}, but struggle with partially synthesized utterances containing both genuine and synthetic segments. In contrast, \ours verifies the physical coupling between speech and ultrasonic articulatory motion. Synthesized segments lack the corresponding live articulation, breaking this cross-modal consistency.

\begin{figure}[t]
\centering
\includegraphics[width=0.7\linewidth]{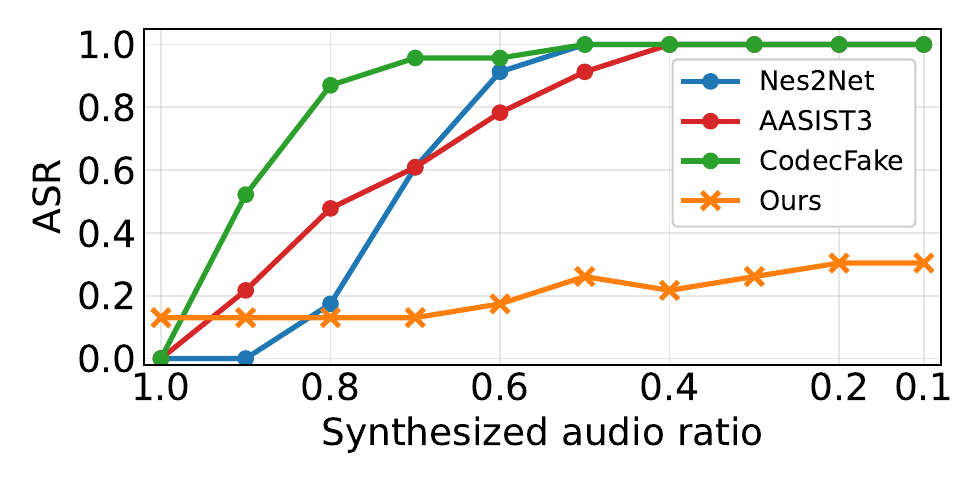}
\caption{ASR under partial synthesis attacks.}
\label{fig:rev:asr_mixed}
\end{figure}

To construct partial synthesis attacks, we start with a genuine 2~s utterance containing synchronized speech and ultrasound. For a synthesized-audio ratio $r\in\{0.1,0.2,\ldots,1.0\}$, we replace a contiguous $2r$-second segment of the audible speech with synthesized audio while keeping the ultrasound unchanged. For example, $r=0.1$ replaces 0.2~s of speech while retaining the remaining 1.8~s of genuine audio. Any utterance with $r>0$ is considered an attack, and ASR measures the fraction incorrectly accepted as genuine. Fig.~\ref{fig:rev:asr_mixed} shows that audio-only methods become increasingly vulnerable as the synthesized portion decreases. While they reliably reject fully synthesized utterances, at $r=0.5$, Nes2Net~\cite{liu2025nes2netlightweightnestedarchitecture} and CodecFake~\cite{xie2024codecfakedatasetcountermeasuresuniversally} reach an ASR of 1.0, while AASIST3~\cite{borodin2024aasist3kanenhancedaasistspeech} reaches 0.91. At $r=0.1$, all three reach an ASR of 1.0. In contrast, \ours maintains substantially lower ASR across synthesized-audio ratios, with ASRs of 0.26 at $r=0.5$ and 0.30 at $r=0.1$. These results demonstrate the benefit of speech--ultrasound coupling for detecting partial synthesis attacks, where synthetic speech cannot reproduce the corresponding live articulatory motion.

\subsection{Real-World Call Robustness}

We evaluate \ours under two practical sources of signal degradation: \textbf{in-call distortions} and \textbf{VoIP codec transcoding}. Overall, \ours remains robust under common distortions and high-fidelity codecs, while aggressive filtering and compression substantially degrade detection.

\noindent\textbf{In-call Distortions.}
Table~\ref{tab:distortion_results} evaluates watermark detection under common in-call distortions. Under clean conditions, \ours achieves an AUC of 1.00, TPR of 0.93, and FPR of 0.01. Performance remains nearly unchanged under resampling, MP3 compression, additive noise, amplitude scaling, smoothing, and low-pass filtering. Echo and Encodec introduce moderate degradation, reducing TPR to 0.80 and 0.58, respectively, while maintaining AUCs above 0.95. More aggressive band-pass and high-pass filtering substantially reduce detection, with AUCs of 0.76 and 0.54. AAC compression is the most challenging distortion, reducing AUC to 0.46 and TPR to 0.01. These results show that \ours tolerates common call distortions but remains sensitive to transformations that substantially alter low-magnitude watermark components.


\noindent\textbf{VoIP Codec Transcoding.}
We further evaluate watermark robustness under representative VoIP codecs, including G.711, G.722, Opus, and Speex. The codec experiments use a separate evaluation pipeline and operating point from the in-call distortion experiments, resulting in different baseline TPR and FPR values. Table~\ref{tab:rev:codec_t2} shows that G.711 and high-bitrate Opus (96~kbps) largely preserve TPR, achieving 0.98 and 0.97, respectively. In contrast, G.722 and lower-bitrate Opus substantially reduce TPR, with Opus at 48~kbps achieving a TPR of 0.08 and FPR of 0.50. Speex at 16~kbps similarly reduces TPR to 0.06. These results demonstrate a trade-off between watermark imperceptibility and robustness to codec compression: \ours remains effective under higher-fidelity codecs, while aggressive compression can substantially weaken the watermark.

\begin{table}[t]
\centering
\footnotesize
\setlength{\tabcolsep}{3pt}
\begin{tabular}{llccc}
\toprule
\textbf{Group} & \textbf{Distortion} & \textbf{AUC} & \textbf{TPR} & \textbf{FPR} \\
\midrule
\textbf{Clean}
& None & 1.00 & 0.93 & 0.01 \\
\midrule
\textbf{Robust}
& Resampling / MP3 & 1.00 & 0.93 & 0.01 \\
& Additive Noise & 1.00 & 0.93 & 0.01 \\
& Amplitude Scaling & 1.00 & 0.93 & 0.01 \\
& Smoothing / Low-pass & 1.00 & 0.92 & 0.01 \\
\midrule
\textbf{Moderate}
& Echo & 0.97 & 0.80 & 0.02 \\
& Encodec & 0.95 & 0.58 & 0.00 \\
\midrule
\textbf{Challenging}
& Band-pass Filter & 0.76 & 0.29 & 0.03 \\
& High-pass Filter & 0.54 & 0.02 & 0.04 \\
& AAC Compression & 0.46 & 0.01 & 0.01 \\
\bottomrule
\end{tabular}
\caption{Robustness under in-call distortions.}
\label{tab:distortion_results}
\end{table}

\begin{table}[t]
\centering
\footnotesize
\begin{tabular}{lcccc}
\toprule
\textbf{Codec} & \textbf{Rate} & \textbf{Bitrate} & \textbf{TPR} & \textbf{FPR} \\
\midrule
Baseline & -- & -- & 0.98 & 0.19 \\
G.711 $\mu$/A-law & 8~kHz & 64~kbps & 0.98 & 0.19 \\
Opus (96 kbps) & 16~kHz & 96~kbps & 0.97 & 0.20 \\
\midrule
G.722 & 16~kHz & 64~kbps & 0.45 & 0.06 \\
Opus (48 kbps) & 16~kHz & 48~kbps & 0.08 & 0.50 \\
Opus (16 kbps) & 16~kHz & 16~kbps & 0.07 & 0.45 \\
Speex & 16~kHz & 16~kbps & 0.06 & 0.01 \\
\bottomrule
\end{tabular}
\caption{Robustness under VoIP codec transcoding.}
\label{tab:rev:codec_t2}
\end{table}

\subsection{Perceptual Quality}

\noindent\textbf{Objective Evaluation.}
We first evaluate whether watermark embedding degrades audio quality. The watermarked speech achieves a PESQ score of 3.99, indicating high perceptual quality. An SI-SNR of 25.6~dB indicates low signal distortion, while a STOI score of 0.922 confirms that speech intelligibility is largely preserved.

\begin{figure}
    \centering
    \includegraphics[width=0.7\columnwidth]{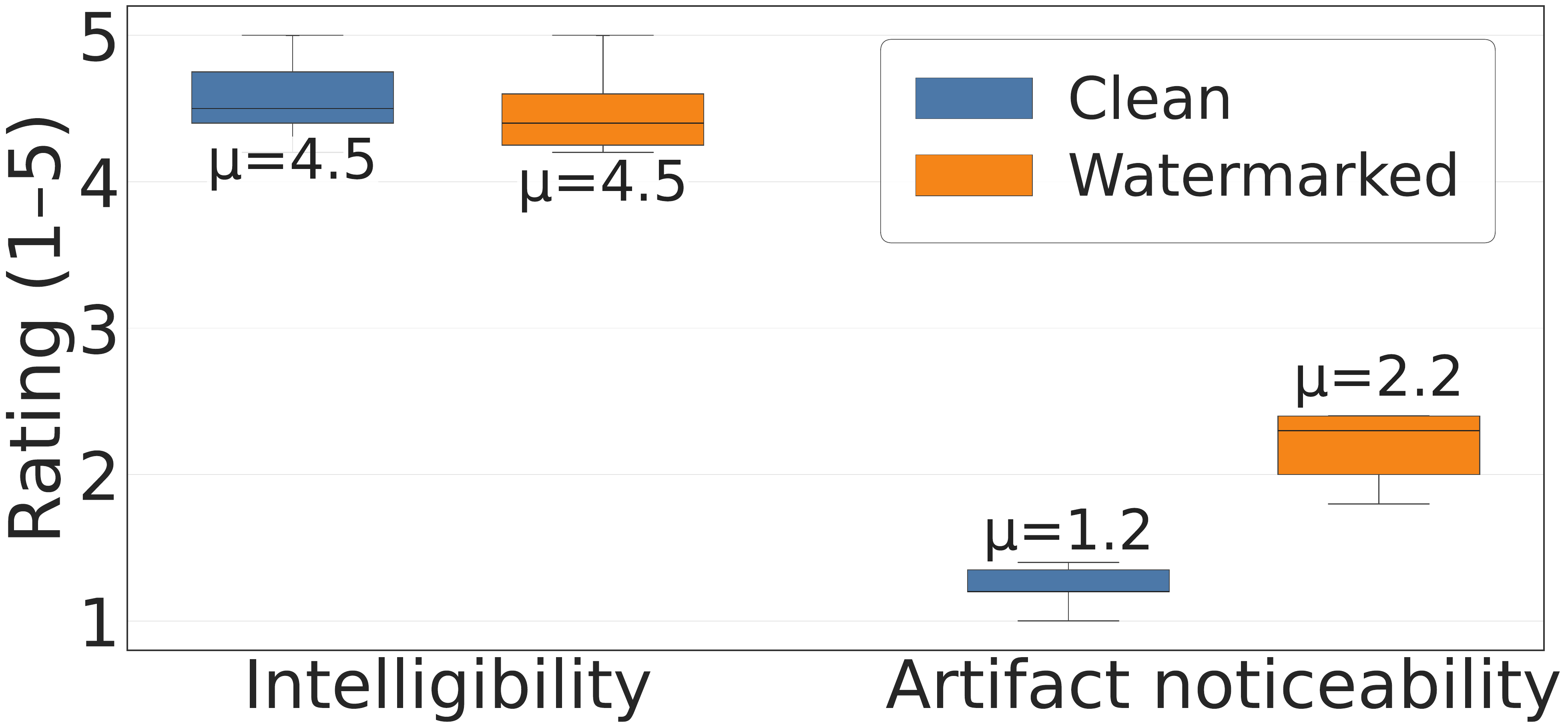}
    \caption{Subjective perception of watermarking.}
    \label{fig:user_study}
\end{figure}

\noindent\textbf{User Study.}
We further conduct a user study to evaluate the perceptual impact of watermarking in a realistic phone-call setting. We use a two-smartphone setup that mimics a live call, with one device generating clean or watermarked speech and the other used for listening. Watermark strength is fixed, and all experiments are conducted on commodity smartphones in quiet indoor environments. Eight participants complete a within-subject study. They listen to sentences with clean and watermarked samples randomly interleaved without knowing the condition. After each sentence, participants rate intelligibility and artifact noticeability on 1--5 Likert scales. As shown in Fig.~\ref{fig:user_study}, intelligibility remains high and comparable between clean and watermarked speech, while artifact noticeability remains low. These results indicate that our system introduces minimal perceptual degradation while preserving speech intelligibility.

\begin{table}[t]
    \centering
    \footnotesize
    \begin{tabular}{lcc}
        \toprule
        \textbf{Environment} & \textbf{Acc.} & \textbf{SNR (dB)} \\
        \midrule
        Office  & 0.97 & 27.46 $\pm$ 6.4 \\
        Outdoor & 0.98 & 26.40 $\pm$ 3.4 \\
        Lab     & 1.00 & 29.15 $\pm$ 6.2 \\
        Home    & 0.95 & 28.99 $\pm$ 5.8 \\
        \midrule
        \textbf{Total} & \textbf{0.97} & \textbf{27.57 $\pm$ 6.1} \\
        \bottomrule
    \end{tabular}
    \caption{Detection performance across environments.}
    \label{tab:env_results}
\end{table}

\subsection{Performance under Natural Variations}

We evaluate whether \ours remains robust under natural variations encountered in everyday phone calls, including different acoustic environments, smartphone placements, call modes, users, and measurement days.

\noindent\textbf{Across Environments.}
We evaluate \ours across four representative environments: office, outdoor, laboratory, and home. These settings capture natural variations in background noise, reverberation, and device placement. As shown in Table~\ref{tab:env_results}, detection accuracy remains above 0.95 in all environments, with an overall accuracy of 0.97. SNR also remains comparable across environments, ranging from 26.40 to 29.15~dB. These results show that \ours maintains consistent detection performance across diverse real-world acoustic environments.

\begin{figure}[t]
  \centering
  \includegraphics[width=\linewidth]{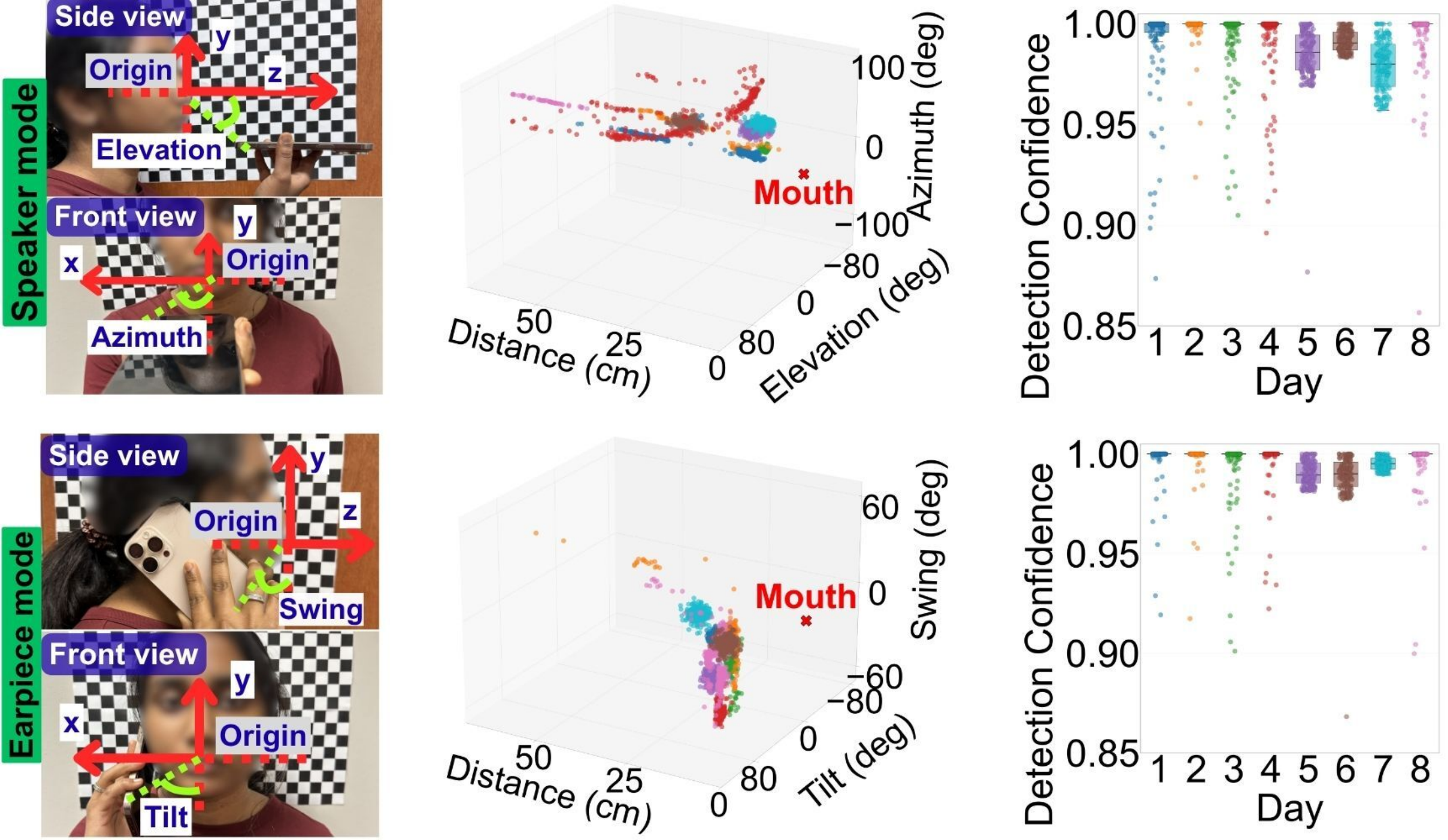}
  \caption{Long-term performance across days and smartphone positions in speaker and earpiece modes.}
  \label{fig:longterm_study}
\end{figure}

\noindent\textbf{Across Days and Smartphone Placements.}
We further evaluate watermark stability over time and under natural variations in smartphone placement through an eight-day study with five participants, each completing one session on a separate day. The end-to-end system runs as native iOS and Android applications, with the generator and detector communicating over a live IP call. During each session, participants stand in front of a calibration checkerboard and naturally hold the smartphone in speaker or earpiece mode, as shown in Fig.~\ref{fig:longterm_study} (left). The Generator app embeds the watermark in real time, while a second device runs the Detector app. Front- and side-view cameras record each session. We define smartphone position as the 3D vector from the mouth center to the phone's bottom-edge center, represented by distance, elevation, and azimuth in speaker mode, and distance, tilt, and swing in earpiece mode.

The watermark is successfully detected across all participants, days, call modes, and observed phone positions. Fig.~\ref{fig:longterm_study} (middle) shows positions sampled every 100~ms and color-coded by day, while Fig.~\ref{fig:longterm_study} (right) shows consistently high batch-wise detection confidence across days and modes. Together with the environmental results, these findings demonstrate robustness to natural variations in acoustic conditions, smartphone placement, user behavior, and time.

\subsection{On-Device Inference Time}
\label{sec:inftime}

Table~\ref{tab:phone_latency} reports the average inference time of the watermark generator and detector across commodity smartphones for each 200~ms audio segment. The generator dominates computation, while the detector incurs relatively low overhead. On recent flagship devices, the generator requires 15--24~ms and the detector less than 8~ms. Even on older devices, total inference time remains below 200~ms per segment, supporting real-time operation on commodity smartphones.

\begin{table}[t]
    \centering
    \footnotesize
    \setlength{\tabcolsep}{4pt}
    \begin{tabular}{lcc}
        \toprule
        \textbf{Phone Model} & \textbf{Generator (ms)} & \textbf{Detector (ms)} \\
        \midrule
        Samsung Z Fold 7       & 19.27  & 1.28 \\
        Samsung Z Flip 7       & 62.35  & 4.32 \\
        Google Pixel 9         & 71.72  & 7.31 \\
        Motorola Edge (2020)   & 160.65 & 21.74 \\
        Motorola Edge 5G UW    & 148.48 & 19.39 \\
        ASUS Zenfone 8         & 134.15 & 10.58 \\
        iPhone 14 Pro Max      & 23.54  & 7.21 \\
        iPhone 16 Pro Max      & 18.41  & 5.88 \\
        iPhone 17 Pro          & 15.28  & 4.72 \\
        \bottomrule
    \end{tabular}
    \caption{On-device inference time across smartphones.}
    \label{tab:phone_latency}
\end{table}
    
\section{Discussion}


\noindent\textbf{Limitations under Double-Watermark Attacks.}
In this attack, an adversary replays the target speaker's already-watermarked speech and applies a second watermarking pass using spoofed ultrasound. Random ultrasound is fully rejected (ASR = 0, AUC = 1.00), whereas speech-synchronized ultrasound achieves an ASR of 0.46 and reduces AUC to 0.89. This exposes a limitation when an attacker can obtain the target's watermarked speech and generate temporally aligned ultrasound. Embedding a fresh, session-specific nonce in each watermark could mitigate this threat.

\noindent\textbf{Limitations under Distortions.}
While \ours is robust to common call distortions and codecs, aggressive spectral filtering and psychoacoustic compression substantially degrade detection. Band-pass and high-pass filtering remove watermark components, while AAC and low-bitrate Opus compression can substantially weaken the watermark through perceptual signal pruning. These results highlight the trade-off between imperceptibility and robustness inherent in low-energy watermarking. Future work will explore multi-band and adaptive embedding strategies.

\noindent\textbf{Limitations under Cross-Speaker Interference.}
The elevated FPRs (0.30--0.37) under speech-synchronized ultrasound attacks (Table~\ref{tab:attack_results}) suggest partial similarity in ultrasonic articulatory patterns across speakers, leading to occasional false acceptance. This indicates a limitation of speaker-agnostic representations. Potential mitigations include speaker-conditioned detection, lightweight speaker embeddings, and more discriminative feature learning.

\section{Conclusion}
We present \ours, a real-time, smartphone-native audio watermarking system that binds speech to synchronized physical articulation through ultrasonic sensing. By jointly leveraging audible speech and ultrasonic articulatory motion, \ours provides a physical basis for verifying speech legitimacy during live calls. Extensive evaluations across diverse users, devices, environments, call conditions, and attack scenarios demonstrate robust detection of replay and synthesized speech while preserving perceptual quality and supporting real-time on-device operation. These results demonstrate the feasibility of using speech--motion coupling as a practical defense against voice impersonation attacks in real-time communication.

\begin{acks}
Generative AI tools were used for language polishing.
\end{acks}
    
    \bibliographystyle{ACM-Reference-Format}
    \bibliography{acmart}
        
    \appendix
    \section*{Appendix}
\label{sec:appendix}

\section{Speech Scripts}
\label{app:data_generation}





Our data collection uses ten sentences representative of common phone-scam scenarios, including urgent account warnings, payment requests, and legal threats, as shown in Fig.~\ref{fig:dataset_scripts}. We use the same scripts for genuine, TTS, and VC speech to maintain consistent linguistic content across attack types. Each sentence is synthesized independently.

\begin{figure}[ht]
\centering
\fbox{
\begin{minipage}{0.9\linewidth}
\textbf{Script 1:} This is an urgent call regarding your account.

\textbf{Script 2:} We have detected suspicious activity on your account.

\textbf{Script 3:} Your account will be suspended if you don't act now.

\textbf{Script 4:} This is Officer Paul from the IRS.

\textbf{Script 5:} You have an outstanding balance that must be paid immediately.

\textbf{Script 6:} Do not hang up or your case will be automatically escalated.

\textbf{Script 7:} We can resolve this issue right now if you follow my instructions.

\textbf{Script 8:} For security reasons, please confirm your full Social Security number.

\textbf{Script 9:} We are calling about a legal matter in your name.

\textbf{Script 10:} There is a warrant out for your arrest.
\end{minipage}
}
\caption{Speech scripts used for data generation.}
\label{fig:dataset_scripts}
\end{figure}

\end{document}